\documentclass{article}

\usepackage{PRIMEarxiv}

\usepackage[utf8]{inputenc} 
\usepackage[T1]{fontenc}    
\usepackage{hyperref}       
\usepackage{url}            
\usepackage{booktabs}       
\usepackage{amsfonts,amsmath}       
\usepackage{nicefrac}       
\usepackage{microtype}      
\usepackage{lipsum}
\usepackage{fancyhdr}       
\usepackage{graphicx}       
\graphicspath{{media/}}     
\newtheorem{definition}{Definition}
\newcommand{\atanTwo}{\operatorname{atan2}}
\usepackage{natbib}
\usepackage[title]{appendix}
\usepackage[symbol]{footmisc}

\title{A Bayesian regression framework for circular models with INLA
}

\author{
  Xiang Ye\textsuperscript{*\dag}, Janet Van Niekerk\textsuperscript{\dag,\ddag}, Håvard Rue\textsuperscript{\dag} \\
  \vspace{5mm} \\ 
  \textsuperscript{\dag} Statistics Program \\
  King Abdullah University of Science and Technology \\
  Thuwal, Saudi Arabia \\
  \texttt{\{xiang.ye, haavard.rue\}@kaust.edu.sa} \\
  \vspace{3mm} \\
  \textsuperscript{\ddag} Department of Statistics \\
  University of Pretoria \\
  Pretoria, South Africa \\
  \texttt{janet.vanniekerk@up.ac.za}
}

\begin{document}
\maketitle

\begin{abstract}
Regression models for circular variables are less developed, since the concept of building a linear predictor from linear combinations of covariates and various random effects, breaks the circular nature of the variable. In this paper, we introduce a new approach to rectify this issue, leading to well-defined regression models for circular responses when the data are concentrated. Our approach extends naturally to joint regression models where we can have several circular and non-circular responses, and allow us to handle a mix of linear covariates, circular covariates and various random effects. Our formulation aligns naturally with the integrated nested Laplace approximation (INLA), which provides fast and accurate Bayesian inference. We illustrate our approach through several simulated and real examples.
\end{abstract}



\section{Introduction}\label{sec1}

Directional statistics, a branch of statistics concerned with data defined on the unit circle in $\mathbb{R}^{2}$ (e.g., wind direction, daily traffic density) or on the hypersphere in $\mathbb{R}^{p}$ for $p \geq 3$ (e.g., detector movement directions, orientations of celestial bodies), has attracted increasing attention over the past decades. For more examples and models in directional statistics, see \citet{mardia2009directional}, \citet{jammalamadaka2001topics}, \citet{ley2017modern}, and \citet{pewsey2021recent}.

One important class of statistical models for studying relationships between variables is the regression model, and a variety of methods in directional statistics have been proposed in recent years to handle regression problems involving circular variables \citep{kim2016regressions}.

Define $\mathcal{S} := \left[-\pi, \pi\right)$ as the circular space represented by radians on the unit circle, where angles that differ by a multiple of $2\pi$ are regarded as equivalent. Let $\mathcal{L}:= \mathbb{R}$ denote the linear space (the real line). Suppose we observe a response variable $y$ of interest, circular variables $\left\{ x_{j}\in \mathcal{S} ; j = 1,\cdots, J \right\}$, and linear variables $\left\{ z_{k}\in \mathcal{L} ; k = 1,\cdots, K \right\}$. Let $\mathcal{C}$ denote the set of covariates included in a given regression model. In general, circular regression can be divided into three categories:
\begin{itemize}
    \item \textbf{circular-linear} ($y \in \mathcal{S}$, $\mathcal{C} \subseteq \left\{ z_{k}\right\}$).
    For example, \citet{george2006semiparametric} model the time of day (on a 24-hour cycle) at which peak ozone concentration occurs, with temperature as a linear covariate.
    \item \textbf{linear-circular} ($y \in \mathcal{L}$, $\mathcal{C} \subseteq \left\{ x_{j}\right\} \cup \left\{ z_{k}\right\}$).
    For example, \citet{alonso2024general} investigate differences between sandhopper species based on their escape directions.
    \item \textbf{circular-circular} ($y \in \mathcal{S}$, $\mathcal{C} \subseteq \left\{ x_{j}\right\} \cup \left\{ z_{k}\right\}$).
    This setting can be viewed as a special case combining circular-linear and linear-circular regression, where circular variables appear both as the model response and as covariates. For example, \citet{kato2008circular} construct a regression model relating wind direction at 6 a.m. to wind direction at noon.
\end{itemize}
Due to the cyclic and non-Euclidean nature of circular data, different regression methods are required for different problems and scenarios. In this section, we first review and discuss circular regression methods available in the literature, and then present the aim of our research.

\subsection{Modeling a circular response}\label{sec1:circular_response}

The common way to construct a linear regression model with a non-Gaussian response is to construct a generalized linear model (GLM) by introducing a link function $g\left(\cdot\right):\mathbb{R}\to \mathcal{D}$ that maps the linear predictor $\eta \in \mathbb{R}$ into the domain $\mathcal{D}$ of the mean (or location parameter) of the response variable. The transformed quantity $g\left(\eta\right)$ is then used to model a distributional parameter of the response. For circular responses, a widely used choice is the inverse tangent link proposed by \citet{fisher1992regression}, given by
\begin{align}
    \label{eq:inverse_tangent_link}
    g\left(z\right) = 2\arctan\left(z\right) \in \left[-\pi,\pi\right), \qquad z \in \mathbb{R}.
\end{align}
An alternative is the scaled probit link given in \citet{mardia2009directional}, of the form $g\left(z\right) = 2\pi\left(\Phi\left(z\right)-0.5\right)$, where $\Phi\left(\cdot\right)$ is the cumulative distribution function of the standard normal distribution.

For a circular variable $x$, most probability density functions for circular distributions include the term $\cos\left(x-\mu\right)$, where $\mu \in \left[-\pi,\pi\right)$ is a location parameter representing the ``mean'' (or dominant) direction of the data. One of the most widely used circular distributions is the von Mises (vM) distribution, also known as the circular normal distribution, which has density
\begin{align}
    \label{eq:vM_density}
    p_{\operatorname{vM}}\left(x \mid \mu, \kappa\right) &= \frac{1}{2\pi \mathcal{I}_{0} \left(\kappa\right)} \exp \left\{ \kappa \cos \left( x - \mu \right) \right\}, \quad \mu \in \left[-\pi,\pi\right), \quad \kappa \in \left[0,\infty\right).
\end{align}
Here, $\kappa$ is a concentration parameter, analogous to the precision parameter of the Gaussian distribution. The density plot for vM distribution with different $\mu$ is given in the left panel of \autoref{fig:compare_dens}.

A \textbf{circular-linear} regression model with a vM response can be constructed through a GLM approach by setting $\mu = g\left(\eta\right)$. An alternative classical circular-linear regression model proposed by \citet{fisher1992regression} assume the mean direction model as follows:
\begin{align}
    \mu_{i} &= \mu_{0} + g\left(\boldsymbol{\beta}^{T}\mathbf{z}_{i}\right),
\end{align}
where $\mu_{0}$ is a baseline direction and $\boldsymbol{\beta}$ are regression coefficients associated with the linear covariates $\mathbf{z}_{i}$ for observation $i$.  Since the intercept is no longer a part of the linear predictor, this formulation is not a GLM; instead, it describes how the linear predictors affect the mean direction of the circular response in terms of geodesic distance. Since the link satisfies $g\left(0\right) = 0$, then $\boldsymbol{\beta}=\boldsymbol{0}$ implies that the linear predictor has no effect on the mean direction.

In both approaches, the difference between $x$ and $\mu$ is not simply given by $x-\mu$. Instead, the distance between two points on the unit circle is defined in terms of a geodesic distance, formalised in Definition~\ref{def:circular_distance}. In other words, the distance between $x$ and $\mu$ is not defined in the linear space.
\begin{definition}[Circular distance] \label{def:circular_distance}
$\forall x_{1},x_{2} \in \mathcal{S}$, the signed geodesic difference between $x_{1}$ and $x_{2}$ is
    \begin{align}
        d_{\mathcal{S}}\left(x_{1},x_{2}\right) := \atanTwo\left( \sin\left(x_{1}-x_{2}\right), \cos\left(x_{1}-x_{2}\right) \right) \in \left[-\pi,\pi\right).
    \end{align}
\end{definition}
Therefore, the link $g\left(\cdot\right)$ is not only mapping the linear predictor to the domain of the response variable, it is actually projecting the linear space onto the circular space, i.e., $g\left(\cdot\right):\mathcal{L}\rightarrow\mathcal{S}$.

However, the both the GLM and non-GLM approaches encounter the same issue. When inference is performed for a single circular variable, the mean direction $\mu \in \left[ -\pi, \pi \right)$ is fixed, and the geodesic difference $x - \mu \in \left[-\pi - \mu, \pi - \mu\right) \Longleftrightarrow x - \mu \in \left[-\pi, \pi \right)$, admits a unique principal representation on the circle. In a regression setting, however, the mean direction $\mu_{i}$ varies across observations. As a result, the corresponding geodesic differences is $x_{i} - \mu_{i} \in \left[\min\left(-\pi - \mu_{i}\right), \max\left(\pi - \mu_{i}\right)\right) \Longleftrightarrow x_{i} - \mu_{i} \in \left[-2\pi, 2\pi \right)$, repeat cyclically across the circle, no longer have a unique representation.

\begin{figure}[!ht]
    \centering
    \begin{minipage}[b]{0.49\textwidth}
        \includegraphics[width=\textwidth]{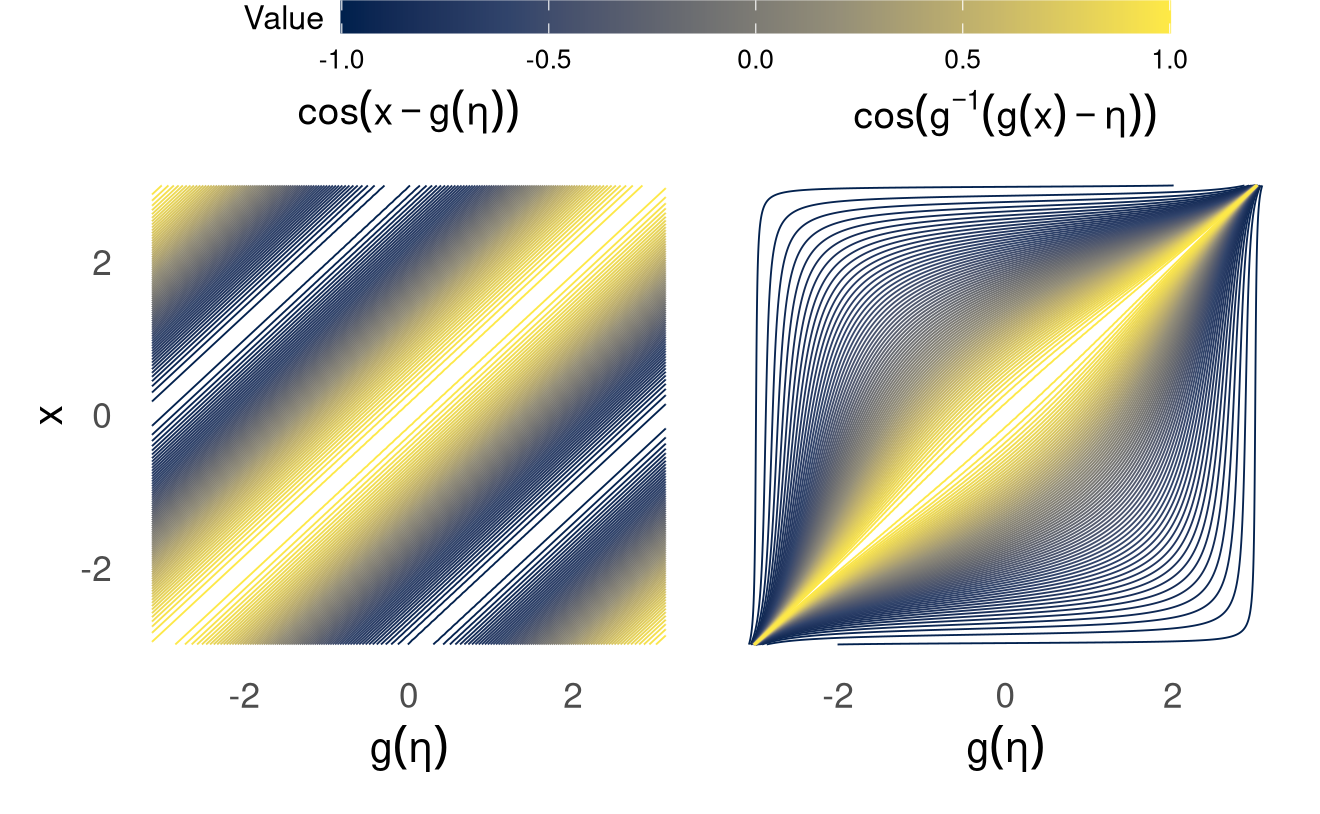}
    \end{minipage}
    \hfill
    \begin{minipage}[b]{0.49\textwidth}
        \includegraphics[width=\textwidth]{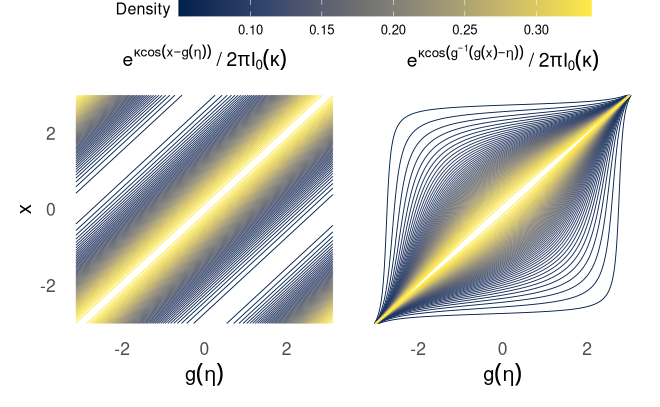}
    \end{minipage}
    \caption{Comparison of circular and linear subtraction between the variable $y$ and the GLM linear predictor. From left to right: (1) cosine values via circular subtraction; (2) cosine values via linear subtraction; (3) von Mises density via circular subtraction; and (4) von Mises density via linear subtraction, when $\kappa=1$.}
    \label{fig:compare_cosine}
\end{figure}

Using the GLM formulation as an illustration, Panel (a) of \autoref{fig:compare_cosine} shows that the cosine term in the density, based on the geodesic difference $x-g\left(\eta\right)$, remains inherently cyclic, exhibiting two repeating modes. Consequently, the corresponding likelihood surface with respect to $\mu$ and $x$ contains two equivalent global optima (Panel (c) of \autoref{fig:compare_cosine}). This multimodality complicates inference in both optimisation-based estimation and simulation-based inference methods, as algorithms may converge to different, equally optimal modes, leading to ambiguous or misleading estimates of the regression parameters.

\subsection{Modeling with circular covariates}\label{sec1:circular_covariates}

When a circular covariate $x$ is present, one modeling approach is to assume that $\cos\left(x\right)$ and $\sin\left(x\right)$ enter the linear predictor with linear effects. Under this assumption, \citet{mardia2009directional} introduced the regression model
\begin{align}
    \label{eq:linear.circular.regression.sincos}
    y_{i} = \beta_{0} + \alpha_{1}\cos x_{i} + \alpha_{2} \sin x_{i} + \epsilon_{i},
\end{align}
where $\alpha_{1}$ and $\alpha_{2}$ are linear coefficients and $\epsilon_{i}$'s are independent Gaussian errors. However, as all information in $x_{i}$ is encoded through $\cos x_{i}$ and $\sin x_{i}$, their induced correlation should be considered \citep{mardia1976linear} and incorporated into the model to obtain more accurate inference.

\citet{alonso2024general} proposed an alternative approach based on circular local likelihood regression. The method allows the effect of a circular covariate to enter the linear predictor through the coefficients of a Taylor-like expansion around a target value $x_{0}$ of the circular variable $x$. For each $x_{0}$, the regression log-likelihood is weighted by a von Mises kernel (\autoref{eq:vM_density}), resulting in a locally weighted, expansion-based approximation to the log-likelihood.

This approach is flexible and accommodates a variety of response families, including Gaussian, Poisson, Gamma and Binomial. It provides local estimates of the effect of the circular covariate at each value $x_{0}$ of interest, allowing functional (nonparametric) effects to be investigated. Extensions to multiple circular and linear covariates are also discussed by the authors. While the method is suited for functional effects analysis, it does not explicitly model structural dependence, such as temporal or spatial correlation, either within the response or among the covariates.

When both the response and the covariates are circular, one approach is to combine existing linear-circular regression methods with the circular-linear regression strategies discussed in Section~\ref{sec1:circular_response}. An alternative is to use a regression model in which some of the coefficients are complex-valued, leading to a fractional linear form. A prominent example is the circular regression model based on the M\"obius transformation on the unit circle, proposed by \citet{kato2008circular}.

\subsection{Multivariate modeling of circular variables}\label{sec1:joint_modeling}

A basic requirement for employing regression models is that the data contain observations of at least two variables. Typically, one variable is regarded as the response variable, and the others are treated as covariates. Regression models provide inference for the effect of covariates on the response. However, the choice of response variable is made by the researcher according to the scientific question of interest. When more than one variable is of substantive interest, one naive approach is to construct several regression models independently.  However, these response variables are often not independent of each other. Treating one response as a covariate in another regression model only captures a one–directional effect and fails to characterise their joint behaviour. To address this, \citet{wulfsohn1997joint} proposed a joint model that simultaneously analyses a survival response and a longitudinal factor through shared random effects instead of including the longitudinal measurement as a covariate in the survival model. A Bayesian formulation of the joint longitudinal-survival framework is given in \citet{henderson2000joint}, and a forthcoming book about this topic is \citet{rustand2026bayesian}.

Joint modeling is becoming increasingly important in directional statistics, where applications frequently involve multiple circular observations. For example, in biomechanical studies, the Ilizarov ring fixator is used in orthopedic surgery to stabilise fractures or severe bone damage. The fixator may exhibit small linear or angular displacements, and understanding its reliability requires modeling dependence across displacements in different directions \citep{nagar2024dependent}. Other examples include studying the 3D structure of proteins \citep{kato2024versatile, hamelryck2025unfolding} and analyzing running gait cycles through joint angles of the hip, knee, and ankle \citep{cohen2024bayesian}.

The main advantage of joint circular regression is that it avoids the multimodality in the likelihood induced by geodesic-distance–based operations when performing addition or subtraction between circular variables, or when combining circular and linear variables. However, modeling dependence in the presence of circular variables remains challenging. Common strategies include using spherical or hyperspherical distributions to represent collections of circular variables, or constructing specialised copulas \citep{nagar2024dependent, kato2024versatile} to accommodate dependence structures among multiple circular variables, as well as between circular and linear variables.

\subsection{Aim of the paper}\label{sec1:aim}

Our aim is to construct a flexible Bayesian regression framework capable of handling regression models that involve several circular and linear variables, with circular variables appearing either as covariates or responses, and allowing for both fixed and random effects.

Since most commonly used responses and covariates are in linear space, regression models are typically formulated within a linear framework. A key assumption of linear regression is that all variables are in $\mathcal{L}$, and it is not well defined to directly combine variables in the circular space $\mathcal{S}$ with variables in the linear space $\mathcal{L}$ through simple addition. Consequently, additional strategies are required when circular variables are involved. In particular, one must address how the circular nature of the response is properly respected and how circular variables enter the linear predictor.

We tackle the challenges arising in both circular-linear and linear-circular regression settings step by step. We first focus on how to connect a circular response to a linear predictor in a way that avoids the inferential ambiguity caused by cyclic likelihood structure. We then turn to the case where circular variables enter the predictor, and demonstrate how their information can be incorporated in a well-defined regression formulation. Finally, we bring these ingredients together in a unified Bayesian framework that borrows strength from multivariate modeling to accommodate a wide range of regression settings.

\subsection{Structure of the remaining part of the paper}

Section~\ref{sec2} introduces the preliminary concepts required for this work, and Section~\ref{sec3} illustrates the proposed circular regression methodology. Simulation studies are presented in Section~\ref{sec4} to demonstrate the stability and flexibility of the proposed framework, followed by two real-world application examples in Section~\ref{sec5}. Finally, conclusions are discussed in Section~\ref{sec6}.


\section{Preliminaries}\label{sec2}

The concept of the latent Gaussian process is discussed in Section~\ref{sec2:latent_gaussian_process}, and the Integrated Nested Laplace Approximation (INLA), a fast numerical approach for Bayesian inference, is briefly introduced in Section~\ref{sec2:inla}.

\subsection{Latent Gaussian process}\label{sec2:latent_gaussian_process}

A latent Gaussian process (LGP) is an unobserved stochastic process $\mathbf{w} = \left(w_{1}, \ldots, w_{n}\right)$, indexed by $i=1,\ldots,n$, assumed to follow a multivariate Gaussian distribution,
\begin{align}
    \mathbf{w} \sim N\left(\boldsymbol{\mu}, \boldsymbol{\Sigma}\right) \Longleftrightarrow \mathbf{w} \sim N\left(\boldsymbol{\mu}, \boldsymbol{Q}^{-1}\right),
\end{align}
where $\boldsymbol{\mu}$ is the mean vector, $\boldsymbol{\Sigma}$ is the covariance matrix, and $\boldsymbol{Q}$ is the precision matrix, which is typically sparse. By choosing different structures for $\boldsymbol{Q}$, the LGP can introduce various forms of conditional dependence into the model, such as temporal correlation (e.g., autoregressive or random walk models) \citep{rue2005gaussian}, spatial dependence, e.g., Matérn SPDE models, \citep{lindgren2011explicit}, or smooth functional effects (e.g., spline-based models) \citep{lang2004bayesian}.

Let $\mathbf{w}$ denote all Gaussian variables, including the LGPs, and let $\boldsymbol{\theta}_{1}$ be the hyperparameters of $\mathbf{w}$. Denote the observational variables by $\mathbf{y}=\left\{y_{i}:i\in\mathcal{I}\right\}$, with likelihood  $p\left(\mathbf{y}\mid\mathbf{w},\boldsymbol{\theta}_{2}\right)$, as commonly assumed in latent Gaussian models, where $\boldsymbol{\theta}_{2}$ are the likelihood hyperparameters. For simplicity, let $\boldsymbol{\theta} = \left(\boldsymbol{\theta}_{1}^{T}, \boldsymbol{\theta}_{2}^{T}\right)^{T}$. The posterior distribution is
\begin{align}
    \label{eq:LGM}
    p\left(\mathbf{w},\boldsymbol{\theta}\mid\mathbf{y}\right) \propto p\left(\boldsymbol{\theta}\right)p\left(\mathbf{w}\mid\boldsymbol{\theta}\right)\prod_{i\in\mathcal{I}}p\left(y_{i}\mid w_{i},\boldsymbol{\theta}\right) = p\left( \mathbf{w},\boldsymbol{\theta}, \mathbf{y} \right).
\end{align}

In Bayesian inference, an LGP serves as a hidden layer that governs or explains how the observed data are generated, allowing complex dependence structures to be incorporated within a coherent probabilistic framework. Moreover, since the linear effects $\left\{\beta_{k}\right\}$ associated with covariates $\left\{z_{k}\right\}$ in a regression model are unobserved quantities, they can also be viewed as latent Gaussian processes. A model that includes LGPs, such as the model in \autoref{eq:LGM}, is known as a latent Gaussian model (LGM).

\subsection{Integrated nested Laplace approximation}\label{sec2:inla}

The Integrated Nested Laplace Approximation (INLA) \citep{rue2009approximate} is a fast alternative to Markov chain Monte Carlo (MCMC) for Bayesian inference in LGMs, where the observations depend on unobserved latent processes $\mathbf{w}$.

Recall the LGM given in \autoref{eq:LGM}, the key idea of INLA is to approximate the posterior marginal $p\left(\boldsymbol{\theta\mid\mathbf{y}}\right)$ by
\begin{align}
    \Tilde{p}\left(\boldsymbol{\theta\mid\mathbf{y}}\right) \propto \frac{p\left( \mathbf{w},\boldsymbol{\theta}, \mathbf{y} \right)}{\Tilde{p}_{G}\left( \mathbf{w} \mid \boldsymbol{\theta}, \mathbf{y} \right)}\Big\vert_{\mathbf{w}=\mathbf{w}^{*}\left(\boldsymbol{\theta}\right)},
\end{align}
where $\Tilde{p}_{G}\left( \mathbf{w} \mid \boldsymbol{\theta}, \mathbf{y} \right)$ is a Gaussian approximation of $p\left( \mathbf{w} \mid \boldsymbol{\theta}, \mathbf{y} \right)$, and $\mathbf{w}^{*}\left(\boldsymbol{\theta}\right)$ is its mode.

Finally, the posterior marginals are approximated via
\begin{align}
    \Tilde{p}\left(w_{i}\mid\mathbf{y}\right) &= \int \Tilde{p}\left(w_{i}\mid \boldsymbol{\theta},\mathbf{y}\right)\Tilde{p}\left(\boldsymbol{\theta}\mid\mathbf{y}\right)d\boldsymbol{\theta}, \\
    \Tilde{p}\left(\theta_{i}\mid\mathbf{y}\right) &= \int \Tilde{p}\left(\boldsymbol{\theta}\mid\mathbf{y}\right)d\boldsymbol{\theta}_{-i},
\end{align}
where $\Tilde{p}\left(w_{i}\mid \boldsymbol{\theta},\mathbf{y}\right)$ can be obtained through a Gaussian approximation, a Laplace approximation, or a simplified Laplace approximation.

Since INLA does not rely on sampling, it avoids the convergence and mixing issues that often arise in MCMC methods. Consequently, it is much faster, as the computations are based on deterministic Laplace approximations and sparse matrix operations. For details on INLA and its developments, see \citet{rue2009approximate}, \citet{lindgren2015bayesian} and \citet{van2023new}.

These features make INLA particularly well suited to joint models, which can be computationally demanding for MCMC methods due to strongly correlated posteriors and slow mixing. In contrast, INLA has been shown to be highly efficient for this class of models \citep{niekerk2021competing, rustand2024joint, rustand2024fast}.


\section{Methodology}\label{sec3}

In this section, we present a Bayesian inference framework that accommodates a wide range of regression models involving circular variables. Our approach to handle circular responses that avoids multimodality in the likelihood is discussed in Section~\ref{sec3:circular_response}. The full regression framework is introduced in Sections~\ref{sec3:latent_gaussian_process} and~\ref{sec3:circular_regression_framework}.

\subsection{Link-adjusted von Mises distribution for circular response}\label{sec3:circular_response}

Following the discussion in Section~\ref{sec1:circular_response},  a central issue in regression models with a circular response and a linear predictor is the definition of the distance between the response and the transformed linear predictor $g\left(\eta\right)$. As shown in the Panel (a) and (b) of \autoref{fig:compare_cosine}, the standard formulation of the regression log-likelihood for circular data results in multiple global optima. To avoid the consequential multimodality-induced inferential ambiguity, we will measure distance in the linear space before mapping them back to the circle,
\begin{align}
    g\left( g^{-1}\left(x_{i}\right) - \eta_{i} \right).
\end{align}
This formulation computes the difference $g^{-1}\left(x_{i}\right) - \eta_{i}$  in linear space $\mathcal{L}$, and then projects the resulting value back onto the circular space $\mathcal{S}$. In this case, the geodesic difference entering the cosine function is restricted to $\left[-\pi,\pi\right)$, matching the natural support of the von Mises distribution. Moreover, as shown in the Panel (c) and (d) of \autoref{fig:compare_cosine}, the cosine component of the likelihood surface under this construction contains only a single global optimum, substantially improving inferential behavior.

However, it is worth noting that this formulation becomes singular at $\left\lvert x \right\rvert = \pi$. Consequently, it is suitable for applications in which the circular response does not occupy the full circle. Particularly, the data should avoid the neighborhood of $\pi$ and $-\pi$. In practice, circular observations can be pre-centered around zero.

\begin{figure}[!ht]
    \centering
    \begin{minipage}[b]{0.49\textwidth}
        \includegraphics[width=\textwidth]{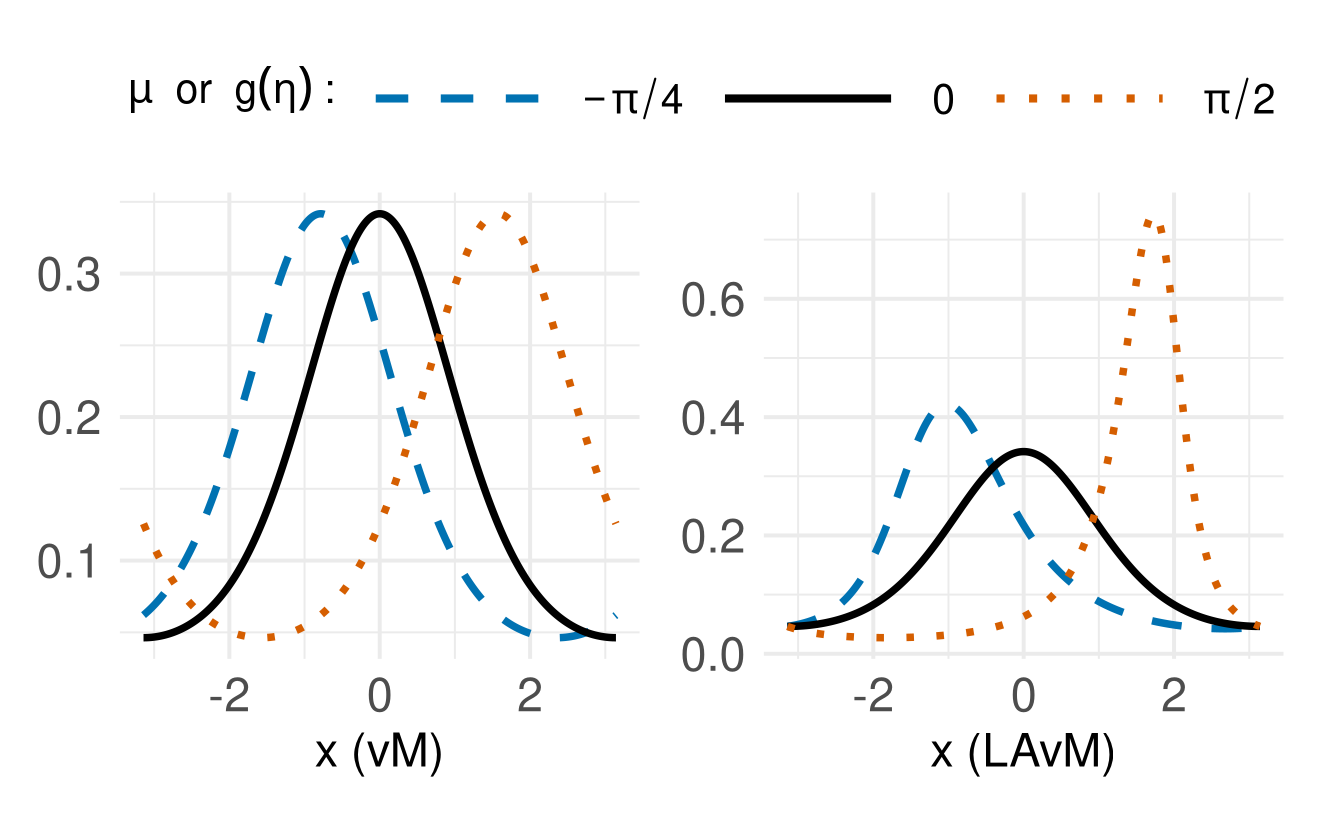}
        \caption{Densities of the vM (left) and LAvM (right) distributions with $\kappa=1$, For the vM distribution, $\mu = -\frac{\pi}{2},0,\frac{\pi}{2}$; for the LAvM distribution, $g\left(\eta\right) = -\frac{\pi}{2},0,\frac{\pi}{2}$.}
    \label{fig:compare_dens}
    \end{minipage}
    \hfill
    \begin{minipage}[b]{0.49\textwidth}
        \includegraphics[width=\textwidth]{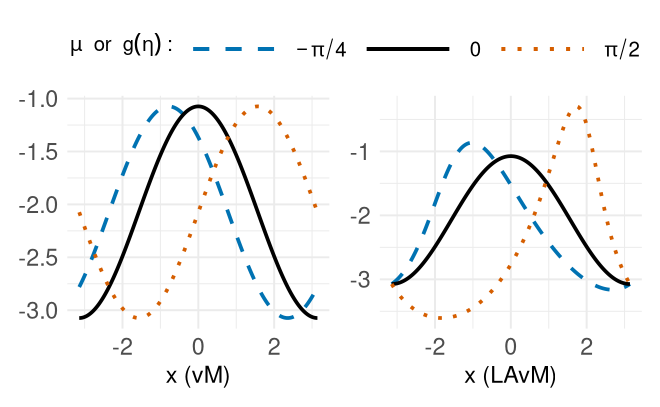}
        \caption{Log densities of the vM (left) and LAvM (right) distributions with $\kappa=1$, For the vM distribution, $\mu = -\frac{\pi}{2},0,\frac{\pi}{2}$; for the LAvM distribution, $g\left(\eta\right) = -\frac{\pi}{2},0,\frac{\pi}{2}$.}
        \label{fig:compare_dens_log}
    \end{minipage}
\end{figure}

The corresponding link-adjusted circular (LAC) distribution is given in Definition~\ref{def:LAC}:
\begin{definition}[Link-adjusted circular distribution]\label{def:LAC}
    Let $z\in\left[-\pi,\pi\right)$ follows a circular distribution with density function $p_{\mathcal{C}}\left(z\mid\mu, \kappa\right)$, with location parameter $\mu\in\left[-\pi,\pi\right)$, and the concentration parameter $\kappa$. Consider a generalized linear model in which the observed circular response variable is $x$ and the linear predictor is $\eta \in \mathbb{R}$. Let the strictly monotone link function be $g\left(\cdot\right): \mathbb{R} \to \mathcal{S}$ with inverse $g^{-1}\left(\cdot\right): \mathcal{S} \to \mathbb{R}$ satisfying $g\left(0\right)=g^{-1}\left(0\right)=0$, then the associated \emph{link-adjusted} distribution for $x$, denoted by $\operatorname{LAC}\left( \eta, \kappa \right)$, has density
    \begin{align}
        p_{\operatorname{LAC}}\left(x \mid \eta, \kappa\right) &= p_{\mathcal{C}}\left(z\mid\mu=0, \kappa\right)\left\lvert \frac{\left(g^{-1}\right)'\left(x\right)}{\left(g^{-1}\right)'\left(z\right)} \right\rvert, \quad x\in\left(-\pi,\pi\right),
    \end{align}
    where $z = g\left( g^{-1}\left(x\right) - \eta \right)$.
\end{definition}
The link-adjusted von Mises (LAvM) distribution associated to the inverse tangent link is found to be
    \begin{align}
        p_{\operatorname{LAvM}}\left(x \mid \eta, \kappa\right) &= \frac{\exp\left\{ \kappa \cos\left( 2 \arctan \left( \tan \left( \frac{x}{2} \right) - \eta \right) \right) \right\}}{2\pi I_{0}\left(\kappa\right)\left(1 + \eta^{2} - \eta \sin\left(x\right) - \eta^{2}\sin^{2}\left(\frac{x}{2}\right)\right)},\quad x\in\left(-\pi,\pi\right).
    \end{align}
In the rest of this paper, we use the inverse tangent link. Some details are given in Appendix~\ref{appendix1}. The mean of the LAvM distribution is attained when $g^{-1}\left(x\right)-\eta=0$, that is, at $x = g\left(\eta\right)$. From the density functions, we observe that when $\eta=0$, the LAvM density equals to the von Mises density. The densities of the vM distribution and the LAvM distribution with inverse tangent link are compared in \autoref{fig:compare_dens}, and the log densities are shown in \autoref{fig:compare_dens_log}. From the plots, we observe that the concentration of the LAvM depends on $\eta$. As $\eta$ moves away from zero, the density becomes increasingly concentrated around $g\left(\eta\right)$. In fact, the approximate concentration parameter of the LAvM distribution is $\kappa\left(1 + \eta^{2}\right)^{2} + \frac{1}{2}\eta^{2}\left(1+\eta^{2}\right)$, or $\kappa\left(1 + \eta^{2}\right)^{2}$ for high values of $\kappa$ (Appendix~\ref{appendix1:proof1_property}).

Since the LAvM distribution is introduced specifically for regression modeling, the variability of the observed circular $x$ is assumed to be driven and explained by the linear predictor. The concentration parameter $\kappa$ in the LAvM distribution plays a role analogous to a precision parameter.

\subsection{Latent Gaussian model for circular response}\label{sec3:latent_gaussian_process}

In most applications, researchers must acknowledge that not all relevant information is observed or recorded. Employing latent Gaussian processes provides a flexible approach to account for such unobserved structure, which is particularly useful when the response variable exhibits dependence, such as temporal or spatial correlation.

When studying a circular variable of interest, the observed data lie in the circular space $\mathcal{S}$. However, underlying structure or dependence can be modeled in the linear space $\mathcal{L}$. For example, consider circular observations $y_{t}$'s collected over time $t=1,\ldots,T$, a corresponding time-series model can be formulated as
\begin{equation}
    \begin{aligned}
        y_{t}\mid w_{t}, a_{0}, a_{1},\kappa &\sim \operatorname{LAvM}\left( a_{0} + a_{1}w_{t} , \kappa \right) \\
        \mathbf{w}                           &\sim N\left( 0, \mathbf{Q}^{-1} \right)
    \end{aligned}
\end{equation}
where $a_{0}$ is an intercept, $a_{1}$ is a scaling parameter, and $\mathbf{w}=\left(w_{1},\ldots,w_{T}\right)^{T}$ is a Gaussian time series model with fixed precision. In this formulation, temporal dependence is not imposed directly on the circular observations $y_{t}$'s through the likelihood. Instead, it is introduced at the latent level through the Gaussian process $\mathbf{w}$. By choosing $\mathbf{w}$ to have an autoregressive (AR) structure, AR-like temporal dependence among the observations is induced through the latent field. If $\mathbf{w}$ is instead chosen to correspond to Gaussian process with a Matérn correlation structure, the model can be extended to study spatial dependence in $y_{t}$'s.

In our approach, circular observations can be explained by suitable latent processes defined in $\mathcal{L}$ allowing dependence structures to be incorporated while respecting the geometry of the circular response.

\subsection{Bayesian joint circular regression framework for circular covariates}\label{sec3:circular_regression_framework}

Special care is required when incorporating a circular covariate into the linear predictor, existing approaches were reviewed in Section~\ref{sec1:circular_covariates}. Following the insight discussed in Section~\ref{sec1:joint_modeling}, an alternative perspective is to regard the circular covariate as an additional response variable of interest and construct a joint model for the circular ``covariate'' and the original response variable.

This formulation is closely related to measurement error models \citep{muff2015bayesian}, in which covariates and responses are modeled jointly to account for uncertainty in the covariates. In our formulation, we treat the circular covariate as an ``imperfectly observed quantity'' and model it through a linear predictor. The resulting linear predictor then carries the information from the circular covariate into the linear predictor of the response.

For simplicity, let $y_{i}$ denote the Gaussian response for observation $i=1,\ldots,N$, let $\left\{z_{ki}\right\}$ denote the values of the linear covariates $\left\{ z_{k}\in \mathcal{L} ; k = 1,\ldots, K \right\}$, and let the $\left\{x_{i}\right\}$s denote the observations of the circular covariate $x$. By treating the circular covariate as an additional response, we can construct a joint model in which the information contained in $x_{i}$ enters the linear predictor of $y_{i}$ indirectly through a shared latent Gaussian process. A basic formulation can be written as
\begin{equation}
    \begin{aligned}
    \label{eq:basic_framework}
        y_{i} \mid w_{i},\boldsymbol{\phi},\boldsymbol{\psi} &\sim N\left( b_{0} + b_{1}\left(a_{0}+a_{1}w_{i}\right) + \sum_{k=1}^{K}\beta_{k}z_{ki} , \tau^{-1} \right), \\
        x_{i} \mid w_{i},\boldsymbol{\phi},\boldsymbol{\psi} &\sim \operatorname{LAvM}\left( a_{0} + a_{1}w_{i} , \kappa \right), \\
        \mathbf{w} &\sim N\left(0, \mathbf{Q}^{-1}\right),
    \end{aligned}
\end{equation}
where $\mathbf{w} = \left(w_{1}, \ldots, w_{N}\right)^{T}$ is a mean zero latent Gaussian process with precision matrix $\mathbf{Q}$, and $\boldsymbol{\phi}$ and $\boldsymbol{\psi}$ represent latent parameters and hyperparameters of the entire model, respectively.

The circular observations $x_{i}$ are explained in the linear space $\mathcal{L}$ by combining a fixed effect $a_{0}$ with the latent process value $w_{i}$. These linear components capture the structure underlying $x_{i}$, and this information is then incorporated into the linear predictor of $y_{i}$ through the term $b_{1}\left(a_{0} + a_{1}w_{i}\right)$, which is the scaled linear predictor, rather than by including $x_{i}$ directly in the regression for $y_{i}$.

This framework is highly flexible. When the effects of the linear covariates on the circular variable are also of interest, the model for $x_{i}$ can be written as
\begin{align}
    \label{eq:cov_in_x}
    x_{i} \mid w_{i},\boldsymbol{\phi},\boldsymbol{\psi} &\sim \operatorname{LAvM}\left( a_{0} + a_{1}w_{i} + \sum_{k=1}^{K}\alpha_{k}z_{ki}, \kappa \right),
\end{align}
where $\boldsymbol{\alpha} = \left( \alpha_{1}, \ldots, \alpha_{K} \right)^{T}$ are additional coefficients. The resulting linear predictor for $x_{i}$ can then be scaled by $b_{1}$ when entering the model for $y_{i}$. Although now $x_{i}$ is no longer explained solely in  $\mathcal{L}$ by the latent process $w_{i}$, the latent process  still carries information about the structure in $x_{i}$ into the linear predictor of $y_{i}$.

Furthermore, when sufficient information or prior knowledge is available, the circular variable $x_{i}$ can be modeled using multiple latent Gaussian processes. For $w_{2t}$ represents temporal dependence at a different time scale from $w_{1t}$:
\begin{align}
    x_{t} \mid w_{1t}, w_{2t},\boldsymbol{\phi},\boldsymbol{\psi} &\sim \operatorname{LAvM}\left( a_{0} + a_{1}w_{1t} + a_{2}w_{2t} , \kappa \right).
\end{align}


\section{Simulation studies}\label{sec4}

In this section, we present three simulation studies to demonstrate our regression framework. The first study is a circular-linear regression with two linear covariates. The second study corresponds to the basic linear-circular regression formulation in \autoref{eq:basic_framework}. Finally, we extent the framework to settings with multiple circular variables and multiple linear response variables. R-code to reproduce this study is available at
\url{https://github.com/XiangYEstats/A-Bayesian-regression-framework-for-circular-models-with-INLA}.

Circular variables are simulated from the LAvM distribution, with sample sizes $N = 100$, $300$ and $1000$. We use the Penalised Complexity (PC) prior for the concentration parameter \citep{ye2025principled}. We denote this prior $\mathcal{PC}_{\kappa}\left(U, \alpha\right)$, where $U$ here indicate the mean resultant length $\rho \in \left[0, 1\right]$ \citep{ley2017modern, lund2017package} defined as
\begin{align}
    \label{eq:mean_resultant_length}
    \rho &= \frac{1}{n}\sqrt{\left( \sum_{i=1}^{n}\cos\left(x_{i}\right) \right)^{2} + \left( \sum_{i=1}^{n}\sin\left(x_{i}\right) \right)^{2}}
\end{align}
for circular data $\left\{x_{i}\right\}$s, and $\alpha \in \left(0,1\right)$. The parameters $U$ and $\alpha$ are chosen following the user's belief of "the probability that the mean resultant length of the data exceeds $U$, is $\alpha$". If there is insufficient prior information, we can chose $U=0.5$ and $\alpha=0.5$.

The other two priors involved in this section are the PC prior for the precision of Gaussian distribution denoted by $\mathcal{PC}_{\tau}\left(U, \alpha\right)$ \citep{simpson2017penalising}, and the PC prior for correlation denoted by $\mathcal{PC}_{cor}\left(U, \alpha\right)$ \citep{sorbye2017penalised}. The parameters are chosen with the prior belief of "the probability that the standard deviation of the data exceeds $U$, is $\alpha$" for $\mathcal{PC}_{\tau}$ prior; And "the probability that the correlation exceeds $U$, is $\alpha$" for $\mathcal{PC}_{cor}$ prior.

Posterior marginals are obtained using R-INLA, with the R package \texttt{INLA} (25.12.16) (see https://www.r-inla.org). Each result is based on averaging over 100 replicates.

To assess the accuracy of the estimates, we report credible intervals of the posterior marginals and posterior predictive distributions of the model response. Posterior means are reported for latent parameters, while posterior modes are reported for hyperparameters. The posterior predictive distribution for a new observation $\tilde{y}$ of the variable $y$ is
\begin{align}
    p\left(\Tilde{y}\mid \mathbf{y}\right) = \int p\left(\tilde{y} \mid \boldsymbol{\theta}\right)\cdot p\left(\boldsymbol{\theta} \mid \mathbf{y}\right)d\boldsymbol{\theta},
\end{align}
where $\mathbf{y}$ denotes the observed data and $p\left(\tilde{y} \mid \boldsymbol{\theta}\right)$ is the sampling model for a new data point conditional on parameter values $\boldsymbol{\theta}$. The posterior predictive distributions is obtained by averaging over 300 simulations.

\subsection{Circular-linear regression} \label{sec4:circular_linear}

\begin{figure}[!ht]
    \centering
    \begin{minipage}[b]{0.5\textwidth}
        \includegraphics[width=\textwidth]{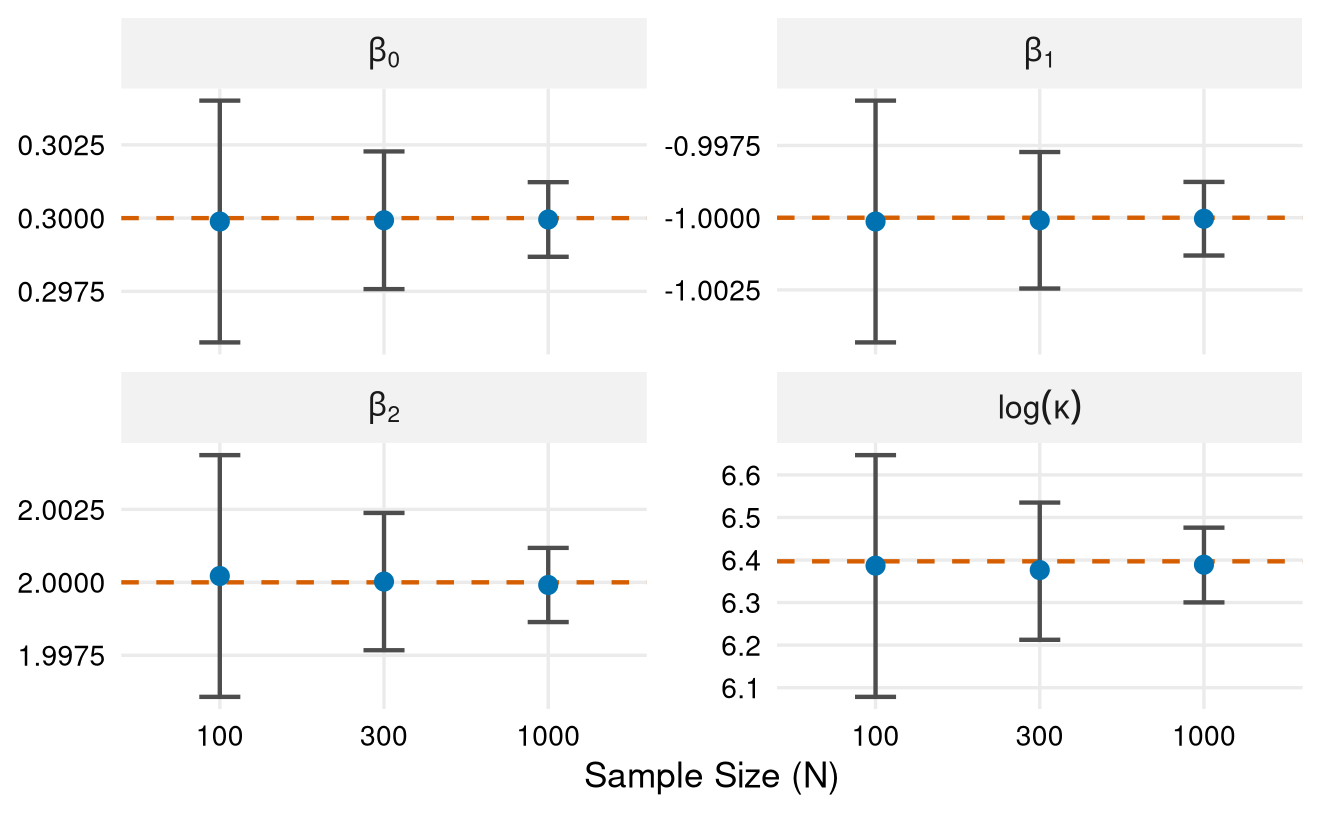}
    \end{minipage}
    \hfill
    \begin{minipage}[b]{0.49\textwidth}
        \includegraphics[width=\textwidth]{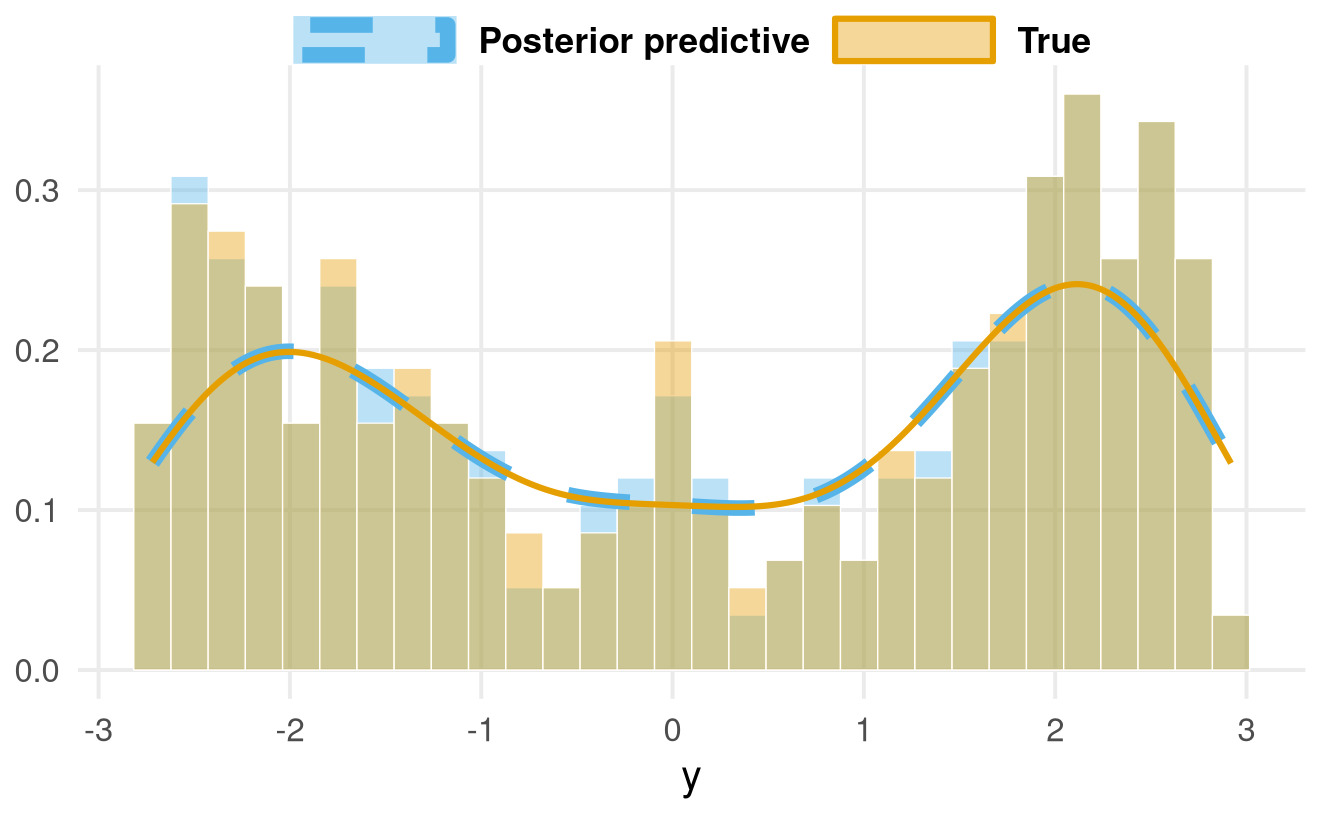}
    \end{minipage}
    \caption{Posterior summaries (left) and posterior predictive distribution (right) for the simulation in Section~\ref{sec4:circular_linear}. In the posterior summaries, the red horizontal lines indicate the true values, the vertical bars represent the 95\% credible intervals, and the blue dots show the posterior means ($\beta_{0}$, $\beta_{1}$, $\beta_{2}$) and mode ($\log\left(\kappa\right)$). In the posterior predictive density plot, the blue histogram and curve correspond to the posterior predictive distribution, while the orange ones denote the true density.}
    \label{fig:sim1}
\end{figure}

The study considers the model
\begin{equation}
    \begin{aligned}
    \label{eq:formula_sim1}
        y_{i} \mid w_{i},\beta_{0},\beta_{1},\beta_{2} &\sim \operatorname{LAvM}\left( \beta_{0} + \beta_{1}z_{1} + \beta_{2}z_{2} , \kappa \right), \\
        \beta_{0},\beta_{1},\beta_{2} &\sim N\left(0, 1\right), \\
        \kappa &\sim \mathcal{PC}_{\kappa}\left(U = 0.5, \alpha = 0.5\right),
    \end{aligned}
\end{equation}
which involves a circular response and two linear covariates. The posterior summaries and posterior predictive distribution for $y$ are shown in \autoref{fig:sim1}. From the posterior summaries, the 95\% credible intervals for all parameters cover the true values. The posterior means of the latent parameters $\beta_{0}$, $\beta_{1}$ and $\beta_{2}$, and the posterior mode of the hyperparameter $\log\left(\kappa\right)$ align well with the true values. The right panel further shows that the posterior predictive distribution closely matches the pattern of the true data.

\subsection{Linear-circular regression} \label{sec4:linear_circular}
This study demonstrate how the proposed framework incorporate circular covariate. The model given in \autoref{eq:formula_sim2} involves one linear variable $y$ and one circular variable $x$:
\begin{equation}
    \begin{aligned}
    \label{eq:formula_sim2}
        y_{i} \mid w_{i},\boldsymbol{\phi},\boldsymbol{\psi} &\sim N\left( b_{0} + b_{1}\left(a_{0} + a_{1}w_{i}\right) , \tau^{-1} \right), \\
        x_{i} \mid w_{i},\boldsymbol{\phi},\boldsymbol{\psi} &\sim \operatorname{LAvM}\left( a_{0} + a_{1}w_{i} , \kappa \right), \\
        \mathbf{w} \sim N\left(0, \mathbf{Q}_{\text{RW2}}^{-1}\right)&, \quad b_{1}, a_{0}, b_{0} \sim N\left(0, 1\right), \\
        \kappa \sim \mathcal{PC}_{\kappa}\left(U = 0.5, \alpha = 0.5\right)&, \qquad 1/a_{1}^{2},\tau \sim \mathcal{PC}_{\tau}\left(U = 0.5, \alpha = 0.5\right).
    \end{aligned}
\end{equation}
The circular variable $x$ goes into the linear predictor for the regression model for $y$ through a latent Gaussian process with RW(2) (second-order random walk) structured precision matrix $\mathbf{Q}_{\text{RW2}}$.

\begin{figure}[!ht]
  \centering
  \begin{minipage}[b]{0.5\textwidth}
    \includegraphics[width=\textwidth]{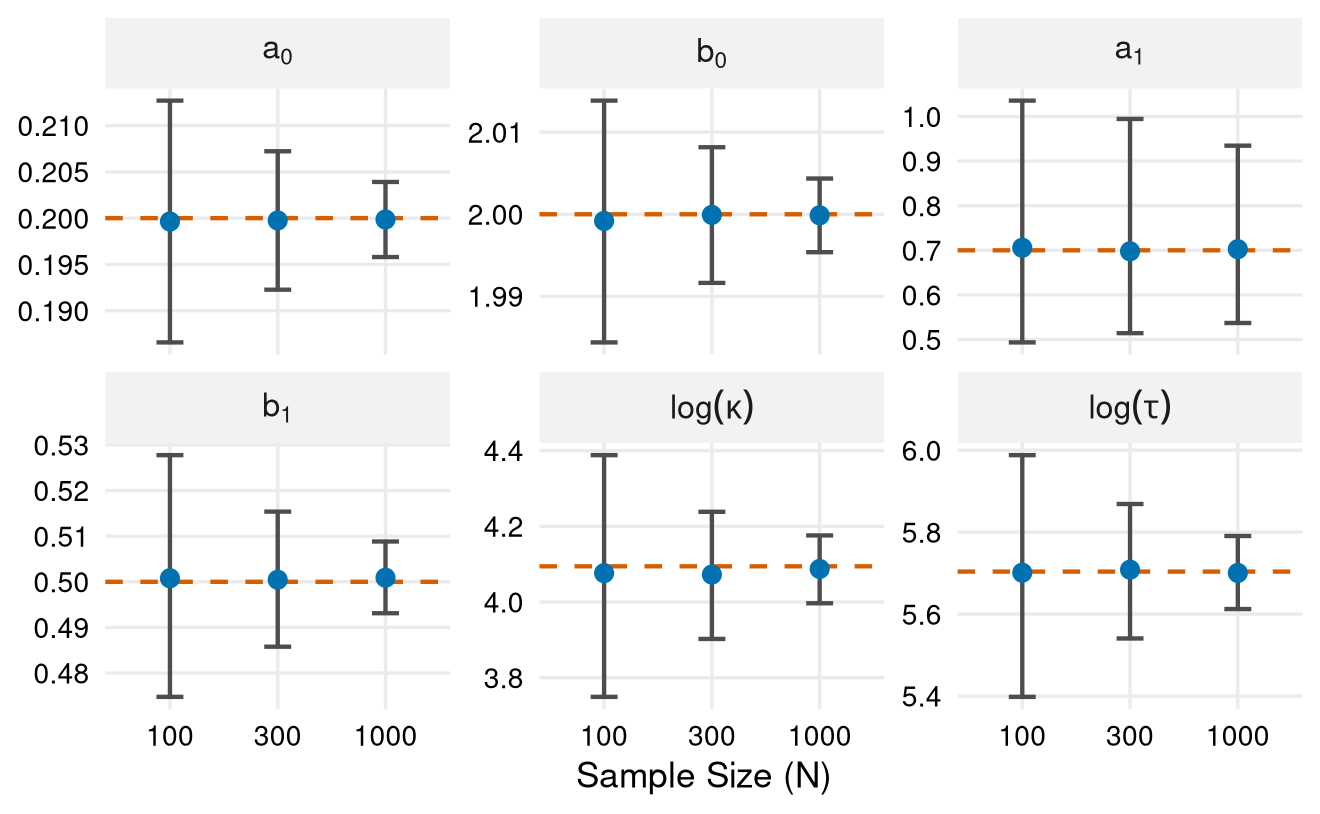}
  \end{minipage}
  \hfill
  \begin{minipage}[b]{0.49\textwidth}
    \includegraphics[width=\textwidth]{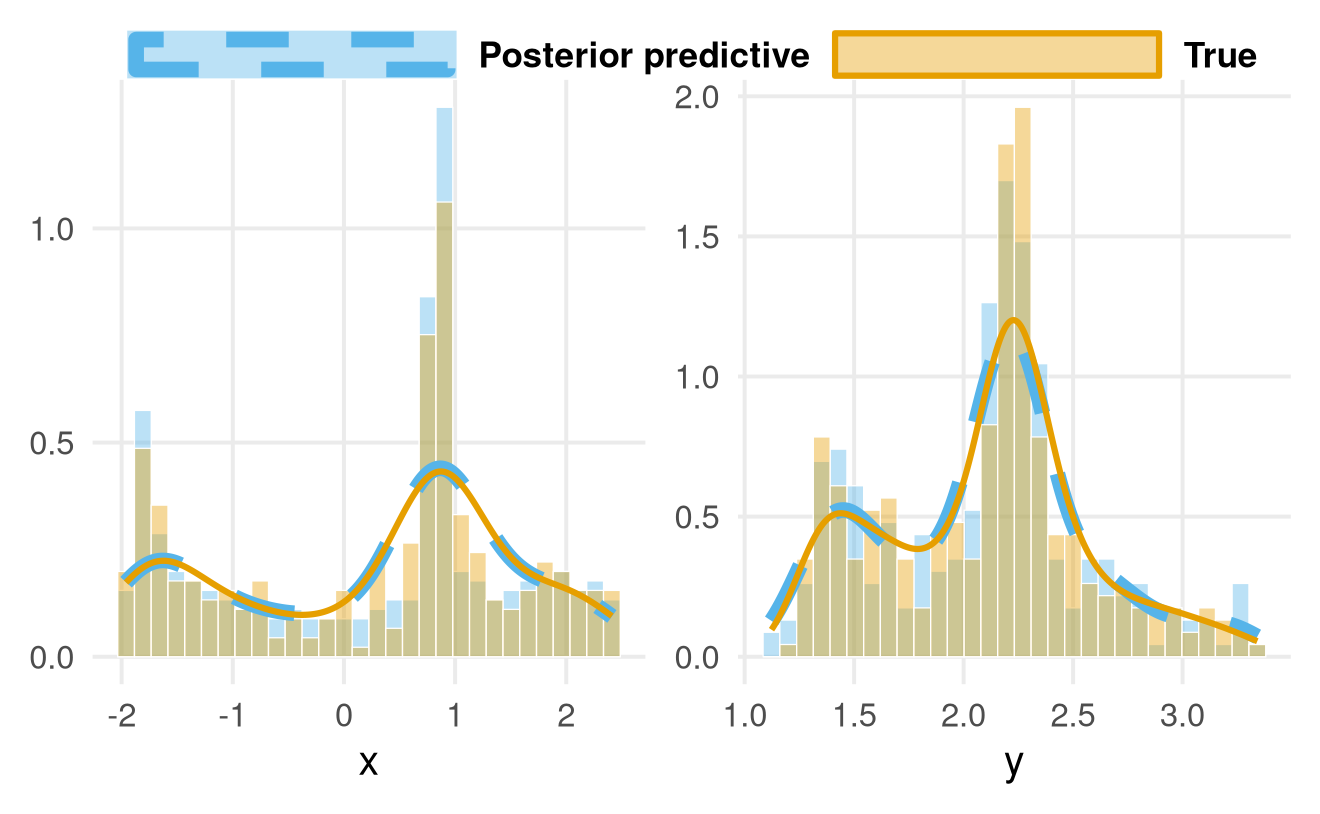}
  \end{minipage}
  \caption{Posterior summaries (left) and posterior predictive distributions (right) for the simulation in Section~\ref{sec4:linear_circular}. The same plotting convention as in \autoref{fig:sim1} is used. The blue dots in the left panel show the posterior means for $a_{0}$, $b_{0}$ and modes for $a_{1}$, $b_{1}$, $\log^{-1}\left(\kappa\right)$, $\log^{-1}\left(\tau\right)$.}
  \label{fig:sim2}
\end{figure}

The posterior summaries and posterior predictive distributions for this simulation study are shown in \autoref{fig:sim2}. From the averaged credible interval of $a_{1}$, we observe that the model captures the scale of the latent process $\mathbf{w}$ well. The credible interval for $b_{1}$ further indicates that the linear predictor associated with $x$ correctly transfers information from $x$ into the linear predictor of $y$. The posterior predictive plots show that the proposed model provides accurate predictions for both $x$ and $y$, and can be used to predict either variable marginally or conditionally on the other.

\subsection{Joint models with random effects and multiple circular variables} \label{sec4:complex_model}

This third study extends the basic setting in \autoref{eq:formula_sim2} to a more general and complex model with two circular variables and two linear response variables. Although this example is stylized, it illustrates the flexibility of the proposed framework. We assume one circular variable can be explained by a second-order random walk latent Gaussian process, and the other is explained by a AR(2) latent Gaussian process. Among the linear responses, one follows a Gaussian distribution ($y_{1}$) and the other follows a Poisson distribution ($y_{2}$). In addition, three linear covariates ($z_{1}$ is Gaussian, $z_{2}$ is Gamma, and $z_{3}$ is Poisson) contribute not only to the linear predictors of the linear responses but also to the circular responses. An i.i.d. random effect ($\mathbf{s}$) is further included in the linear predictor of the Poisson response. The full model is given by
\begin{equation}
    \begin{aligned}
    \label{eq:formula_sim3}
        y_{1i} \mid w_{1i},w_{2i},\boldsymbol{\phi},\boldsymbol{\psi} &\sim N\left( b_{10} + b_{11}\eta_{1} + b_{12}\eta_{2} + \sum_{j=1}^{3}\beta_{1j}z_{ji}, \tau^{-1} \right), \\
        y_{2i} \mid w_{1i},w_{2i}, s_{i}, \boldsymbol{\phi},\boldsymbol{\psi} &\sim \operatorname{Poisson}\left( \exp\left\{ b_{20} + b_{21}\eta_{1} + b_{22}\eta_{2} + \sum_{j=1}^{3}\beta_{2j}z_{ji} + s_{i} \right\} \right), \\
        x_{1i} \mid w_{1i},w_{2i},\boldsymbol{\phi},\boldsymbol{\psi} &\sim \operatorname{LAvM}\left( \eta_{1} = a_{10} + a_{11}w_{1i} + \sum_{j=1}^{3}\alpha_{1j}z_{ji}, \kappa_{1} \right), \\
        x_{2i} \mid w_{1i},w_{2i},\boldsymbol{\phi},\boldsymbol{\psi} &\sim \operatorname{LAvM}\left( \eta_{2} = a_{20} + a_{21}w_{2i} + \sum_{j=1}^{3}\alpha_{2j}z_{ji} , \kappa_{2} \right), \\
        \mathbf{w}_{1} \sim N\left(0, \mathbf{Q}_{\text{RW2}}^{-1}\right)&, \qquad \mathbf{w}_{2} \sim N\left(0, \mathbf{Q}_{\text{AR2}}^{-1}\right), \qquad \mathbf{s} \sim N\left(0,\sigma_{s}\mathbf{I}\right), \\
        b_{11}, b_{12}, b_{21}, b_{22} &\sim N\left(0,1\right), \qquad a_{10}, a_{20}, b_{10}, b_{20}, \boldsymbol{\beta}, \boldsymbol{\alpha} \sim N\left(0, 0.001^{2}\right), \\
        \kappa_{1}, \kappa_{2} \sim \mathcal{PC}_{\kappa}&\left(0.5,  0.5\right), \qquad 1/a_{11}^{2}, 1/a_{21}^{2}, 1/\sigma_{s}^{2}, \tau \sim \mathcal{PC}_{\tau}\left(0.5, 0.5\right), \\
        \text{PACF}\left(1\right), & \text{PACF}\left(2\right) \sim \mathcal{PC}_{cor}\left(0.5,0.5\right).
    \end{aligned}
\end{equation}
Here, the first circular variable $x_{1}$ is driven by a second-order random walk latent Gaussian process, while the second circular variable $x_{2}$ is driven by an AR(2) latent Gaussian process. $\text{PACF}\left(1\right)$ and $\text{PACF}\left(2\right)$ are the Partial Autocorrelation Function (PACF) at Lag 1 and 2, respectively. The PC prior for correlation is used for the hyperparameters PACF(1) and PACF(2) for the AR(2) latent Gaussian process $\mathbf{Q}_{\text{AR2}}$. The two circular variables influence both linear responses through their respective linear predictors. Conditional on the latent field and covariates, the two linear response variables are independent.

Due to the large amount of parameters, only partial simulation results are shown in \autoref{fig:sim3_partial}. The complete posterior summaries for all paramaters are provided in Appendix~\ref{appendix2}. From the first row of the left panel, we observe that the posterior estimates for the coefficients of the linear covariates in the linear predictors of the circular variables ($\alpha_{11}$, $\alpha_{13}$, $\alpha_{22}$) successfully recover the true values, indicating that the effects of the covariates on the circular responses are correctly estimated. In addition, the estimates of the standard deviation of the i.i.d. random effect $\mathbf{s}$ and the PACF parameters of the AR(2) latent Gaussian process show that both the scale and dependence structure of the latent processes are well captured.
\begin{figure}[!ht]
  \centering
  \begin{minipage}[b]{0.5\textwidth}
    \includegraphics[width=\textwidth]{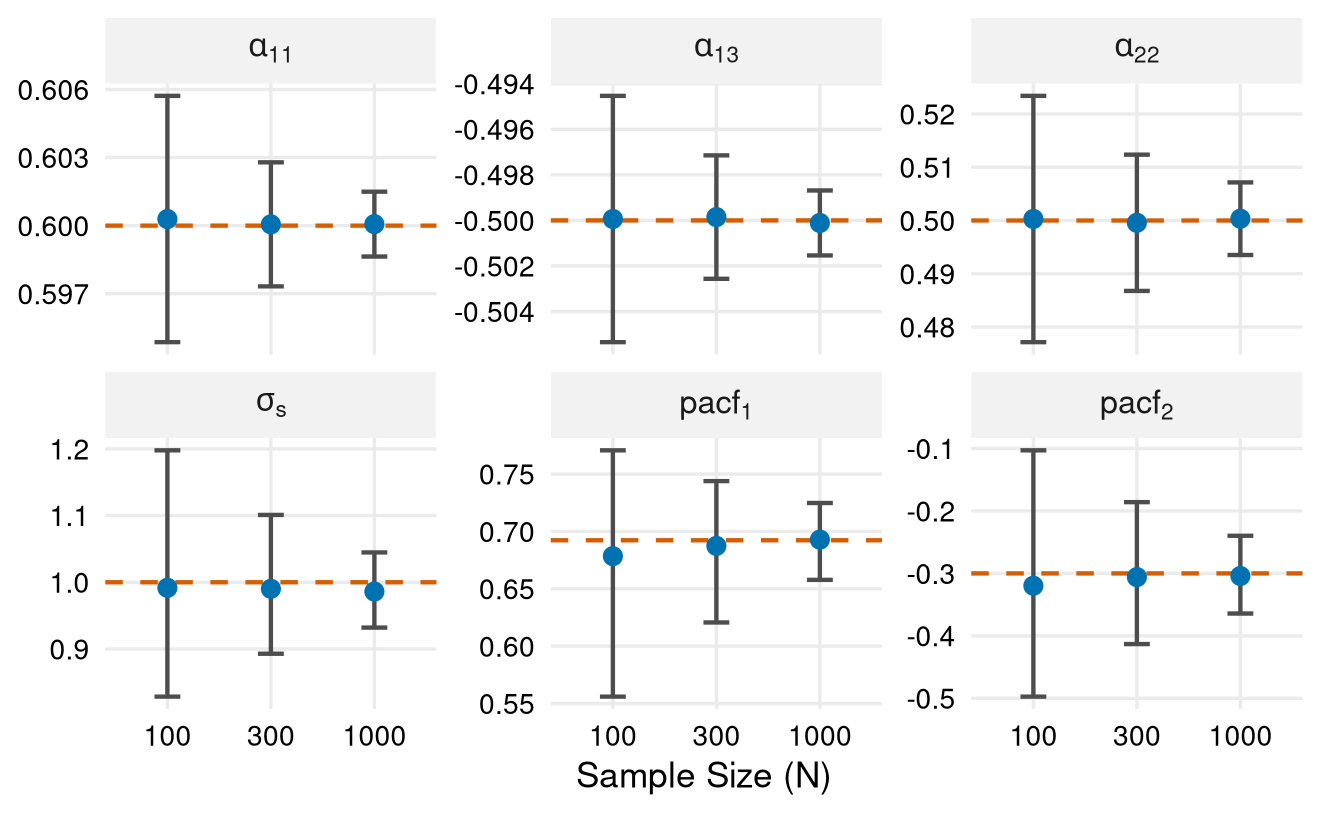}
  \end{minipage}
  \hfill
  \begin{minipage}[b]{0.49\textwidth}
    \includegraphics[width=\textwidth]{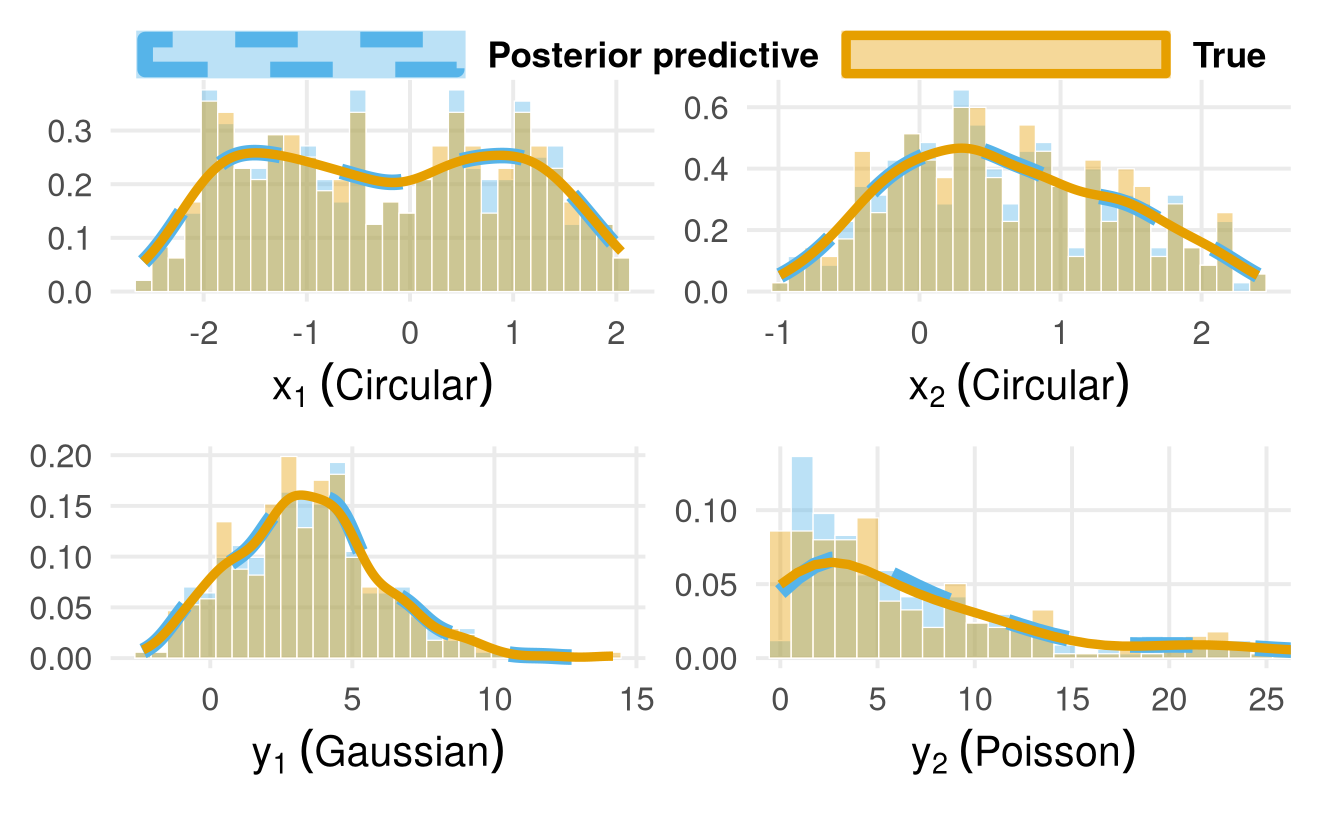}
  \end{minipage}
    \caption{Partial posterior summaries (left) and posterior predictive distributions (right) for the simulation in Section~\ref{sec4:complex_model}. The same plotting convention as in \autoref{fig:sim1} is used. The blue dots in the left panel show the posterior means for $\alpha_{11}$, $\alpha_{13}$, $\alpha_{22}$ and modes for $\sigma_{s}$, $\text{PACF}\left(1\right)$, $\text{PACF}\left(2\right)$.}
    \label{fig:sim3_partial}
\end{figure}

\subsection{Computational time}

All simulations in this section were performed on a Dell XPS 13 9350 workstation running Fedora Linux 43 (Kernel 6.18.5). The machine is equipped with an Intel Core Ultra 7 258V processor (8 cores, single-threaded, up to 4.8 GHz) and 32 GB of RAM.

Among the three simulation studies, the circular–linear model in Section~\ref{sec4:circular_linear} is the fastest, with an average runtime of approximately 1 second, even for sample sizes up to $N=1000$. The model in Section~\ref{sec4:linear_circular}, which incorporates information from a circular variable into the linear predictor of a linear response, requires slightly more computation but remains highly efficient, with an average runtime of about 2.4 seconds for $N=1000$. The model in Section~\ref{sec4:complex_model} has the highest computational cost due to its increased dimensionality, nevertheless, it remains computationally efficient for a multivariate Bayesian joint model, with average runtimes of about 3.2, 6.9, and 18.5 seconds for $N=100$, $300$, and $1000$, respectively.

These computational times show how efficient, yet powerful, the framework is for modeling with circular variables, even for large data sets or complex models.


\section{Application examples}\label{sec5}

We will now illustrate how our framework can be applied to two real data examples.

\subsection{Wind data with temporal dependence}\label{sec5:wind}

The first example aims to study how time, wind direction, and temperature affect wind speed, while simultaneously examining the effects of time and temperature on wind direction. The hourly wind data are collected by the National Centers for Environmental Information (NOAA) \citep{noaa_hourly_normals_2025}. We use data from the first season of 2010 (January, February, and March; a total of 2,159 hourly observations) recorded at the measurement station at John F. Kennedy International Airport in New York, USA. The dataset contains hourly averaged measurements of wind speed ($y$), wind direction ($x$) and temperature ($z$).

Second-order random walk (RW2) and autoregressive (AR2) temporal effects are introduced for wind direction and wind speed through latent Gaussian processes, and the influence of temperature on both variables is explicitly modeled. The model of interest is given by
\begin{equation}
    \label{eq:model_wind}
    \begin{aligned}
        y_{i}\mid w_{i}, s_{i}, \boldsymbol{\phi},\boldsymbol{\psi}  &\sim \operatorname{Gamma}\left( a = \rho, b = \frac{\rho}{\exp\left\{ \eta^{y}_{i}  \right\}} \right) \\
        x_{i}\mid w_{i}, \boldsymbol{\phi},\boldsymbol{\psi} &\sim \operatorname{LAvM}\left( a_{0} + a_{1}w_{i} + a_{2}w_{2i} + \alpha z_{i}, \kappa \right) \\
        \mathbf{w} \sim N\left( 0, \mathbf{Q}_{\text{AR2}}^{-1} \right), \mathbf{w}_{2} \sim &N\left( 0, \mathbf{Q}_{\text{RW2}}^{-1} \right), \mathbf{s}\sim N\left( 0, \sigma_{s}\mathbf{Q}_{\text{AR2}}^{-1} \right), \mathbf{s}_{2}\sim N\left( 0, \sigma_{s_{2}}\mathbf{Q}_{\text{RW2}}^{-1} \right), \\
        b_{1}, a_{0}, &b_{0}, \alpha, \beta \sim N\left(0, 1\right), \quad \kappa \sim \mathcal{PC}_{\kappa}\left(0.5,0.99\right), \\
        1/a_{1}^{2}, 1/a_{2}^{2}, 1/\sigma_{s}^{2},& 1/\sigma_{s_{2}}^{2} \sim \mathcal{PC}_{\tau}\left(0.5,0.5\right), \quad \rho \sim \operatorname{LogGamma}\left(1,0.01\right), \\
        \text{PACF}(1)_{\mathbf{w}},\text{PACF}&(2)_{\mathbf{w}},\text{PACF}(1)_{\mathbf{s}},\text{PACF}(2)_{\mathbf{s}} \sim \mathcal{PC}_{cor}\left(0.5,0.5\right), \\
        \text{where} \quad \eta^{y}_{i} = &b_{0} + b_{1}\left(a_{0} + a_{1}w_{i} + a_{2}w_{2i} + \alpha z_{i}\right) + s_{i} + s_{2i} + \beta z_{i}.
    \end{aligned}
\end{equation}
Here, $\mathbf{Q}_{\text{AR2}}$ a $n \times n$ AR(2) structured precision matrix, where $n$ is the dimension of the data; And $\mathbf{Q}_{\text{RW2}}$ is a cyclic second-order random walk precision matrix defined over a 24-hour cycle. This construction allows the latent process to capture smooth daily patterns that repeat every 24 hours, while ensuring continuity between the end and the beginning of each day \citep[see][Chapter~3]{rue2005gaussian}. The Gamma density is parameterised following the convention used in the \texttt{INLA} package, where $a=\rho$ and $b = \rho/\exp\left(\mu\right)$. The precision parameter $\rho$ is defined by $\tau = \rho/\operatorname{E}\left(y\right)^{2}$ where $\tau = 1/\operatorname{Var}\left(y\right) = b^{2}/a$ denotes the precision of the Gamma distribution.

\begin{figure}[!ht]
  \centering
  \begin{minipage}[b]{0.5\textwidth}
    \includegraphics[width=\textwidth]{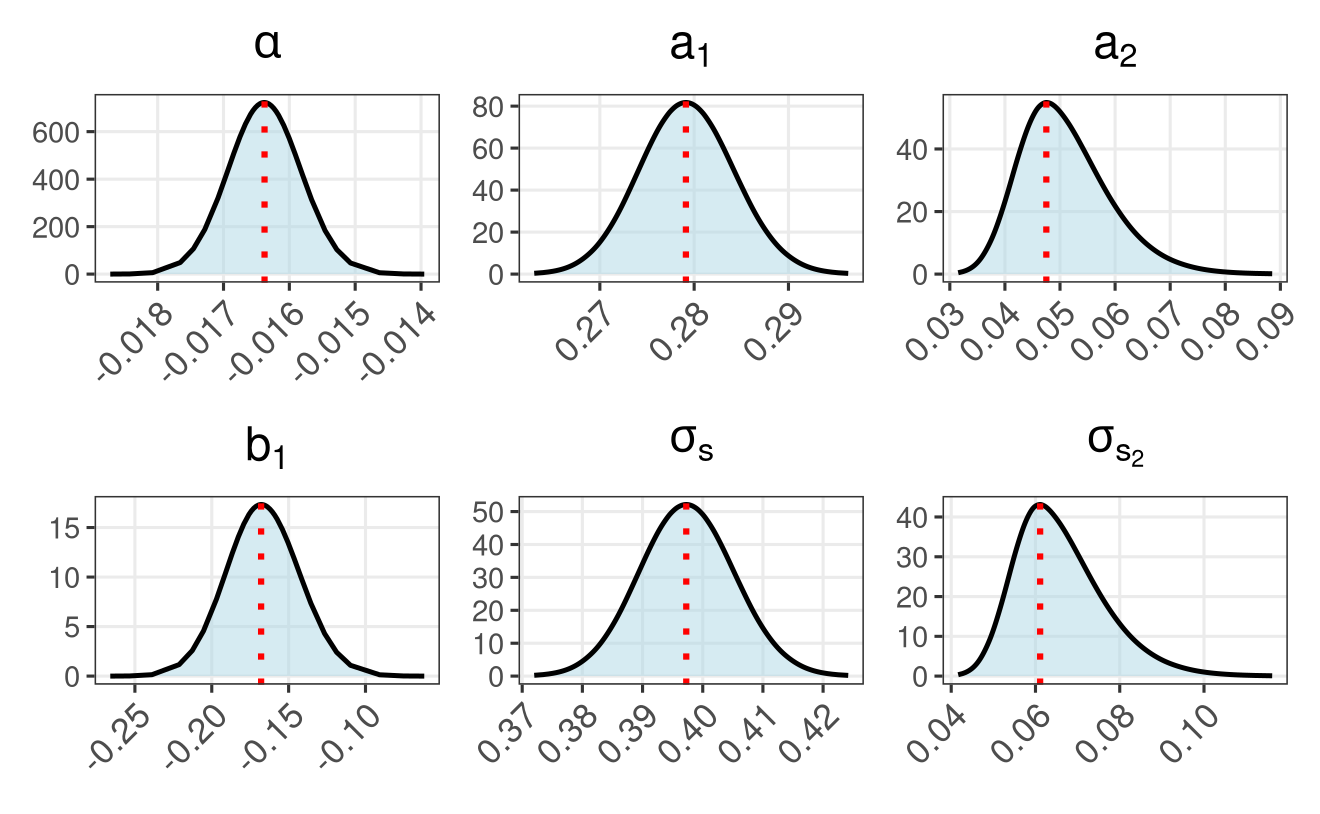}
  \end{minipage}
  \hfill
  \begin{minipage}[b]{0.49\textwidth}
    \includegraphics[width=\textwidth]{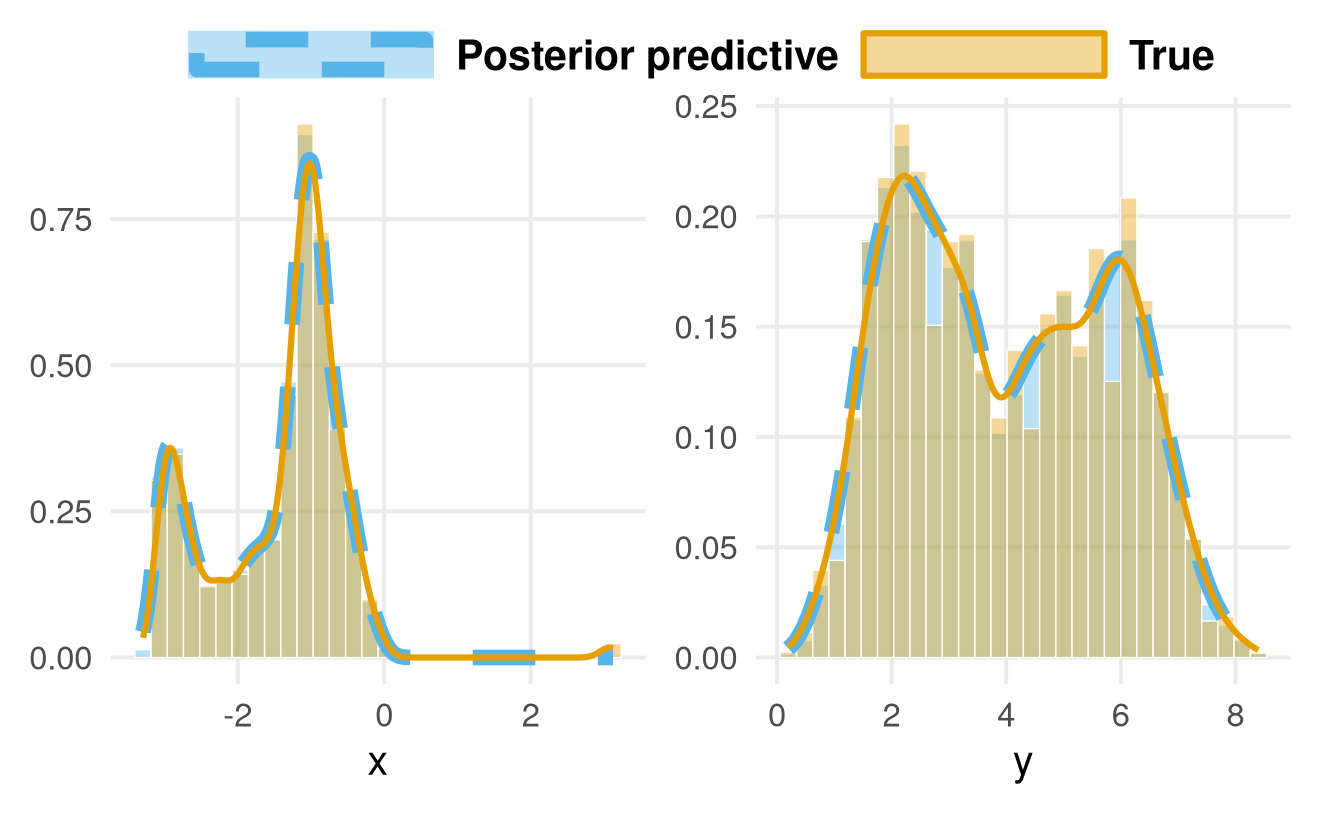}
  \end{minipage}
  \caption{Partial posterior marginals (left) and posterior predictive distributions (right) for the model in \autoref{eq:model_wind}. In the left panel, black curves show posterior densities and red vertical lines indicate posterior means for $\alpha$ and $\beta$, and posterior modes for the remaining parameters. The posterior predictive plots follow the same convention as in \autoref{fig:sim1}.}
  \label{fig:wind}
\end{figure}

The model running time is around 80 seconds, and the posterior marginals for $\alpha$, $a_{1}$, $a_{2}$, $b_{1}$, $\sigma_{s}$ and $\sigma_{s_{2}}$, and the posterior predictive distributions of $x$ and $y$ are given in \autoref{fig:wind}. The posterior marginals for all parameters are presented in Appendix~\ref{appendix2}. From the left panel of the figure, we can observe that the effects of temperature on wind direction ($\alpha$) is significant, with mean around -0.016. The AR(2) and RW(2) latent Gaussian processes for wind direction and wind speed have standard deviation with modes around $a_{1} = 0.279$, $a_{2} = 0.049$, $\sigma_{s} = 0.397$ and $\sigma_{s_{2}} = 0.062$. The effect of wind direction on the wind speed $b_{1}$ is negative, with mode around -0.168.

For comparison, we consider a classical linear-circular regression model in which $\sin\left(x\right)$ and $\cos\left(x\right)$ are included as covariates in the linear predictor for $y$. Model performances are assessed using leave-one-out cross-validation information criterion (LOOIC) \citep{vehtari2017practical} and Geometric Mean of the Conditional Predictive Ordinate (GM-CPO) \citep{gelfand1994bayesian} through leave-one-out cross-validation. The lower LOOIC and higher GM-CPO suggest a better model. The validation results are presented in \autoref{tab:metrics_comparison}.
\begin{table}[!ht]
\centering
\caption{Performance Metrics Comparison: Classical Approach vs. Proposed Framework.}
\label{tab:metrics_comparison}
\begin{tabular}{l c c c c}
\toprule
& Classical Approach & \multicolumn{3}{c}{Proposed Framework} \\
\cmidrule(lr){2-2} \cmidrule(lr){3-5}
Metric & Speed ($y$) & Speed ($y$) & Direction ($x$) & Joint ($y$ \& $x$) \\
\midrule
LOOIC  & -1276.79 & \textbf{-1353.88} & -8563.63 & -8567.13 \\
GM-CPO &     1.34 &     \textbf{1.37} &     7.26 &     7.27 \\
\bottomrule
\end{tabular}
\end{table}

The table shows that the proposed framework gives more accurate predictions of $y$ than the classical approach. In addition, it also provides precise estimates for the circular variable $x$, which the classical model does not attempt to predict. Moreover, the proposed framework also allows joint leave-one-out cross-validation, where each $y_{i}$ is left out together with its corresponding $x_{i}$ \citep{liu2025leave}. This highlights a practical advantage of the proposed joint circular regression approach: information is shared through the latent structure, allowing parallel prediction of both linear and circular components rather than treating the circular observations only as fixed covariate values.

\begin{figure}[!ht]
  \centering
  \begin{minipage}[b]{0.50\textwidth}
    \includegraphics[width=\textwidth]{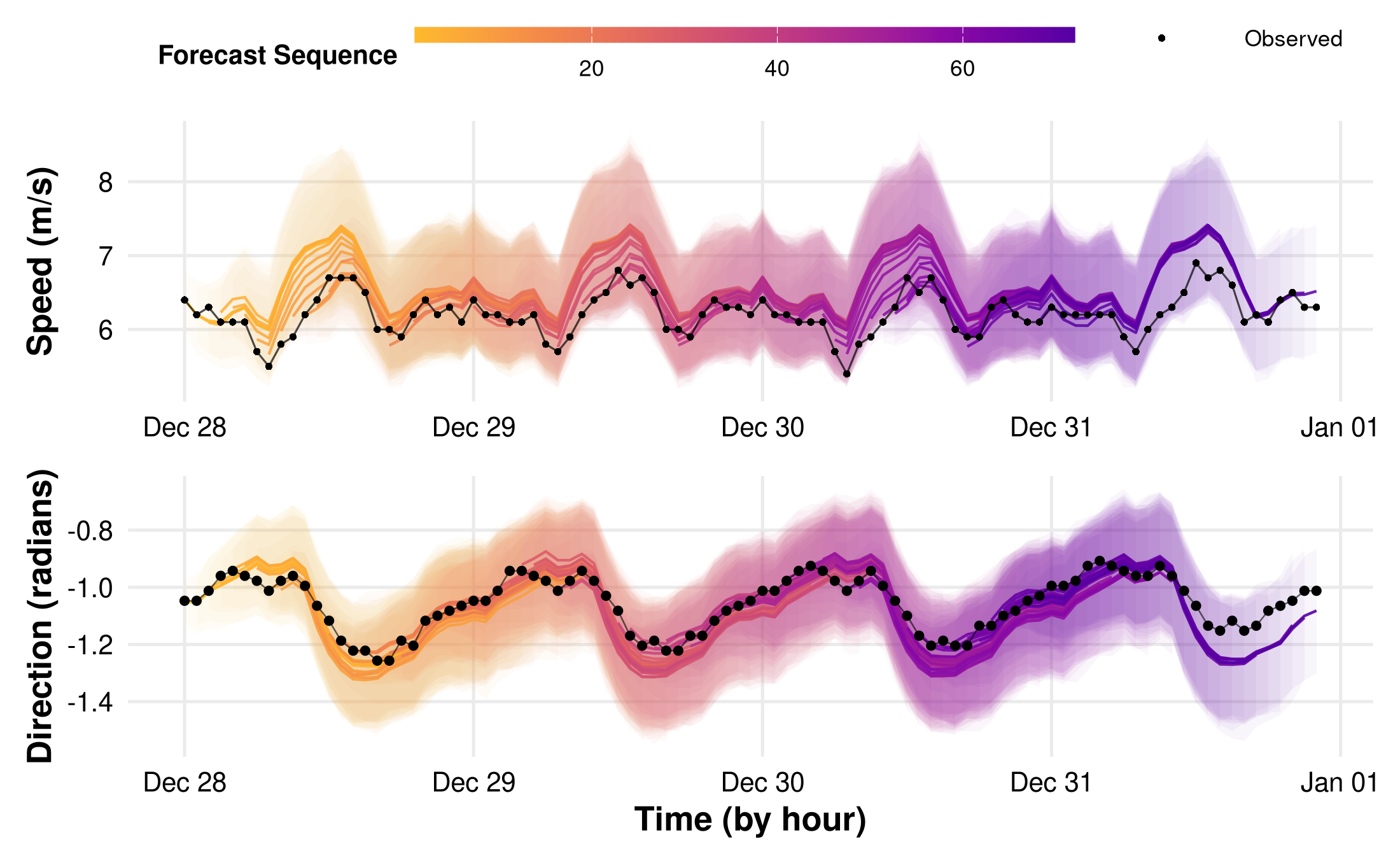}
  \end{minipage}
  \hfill
  \begin{minipage}[b]{0.49\textwidth}
    \includegraphics[width=\textwidth]{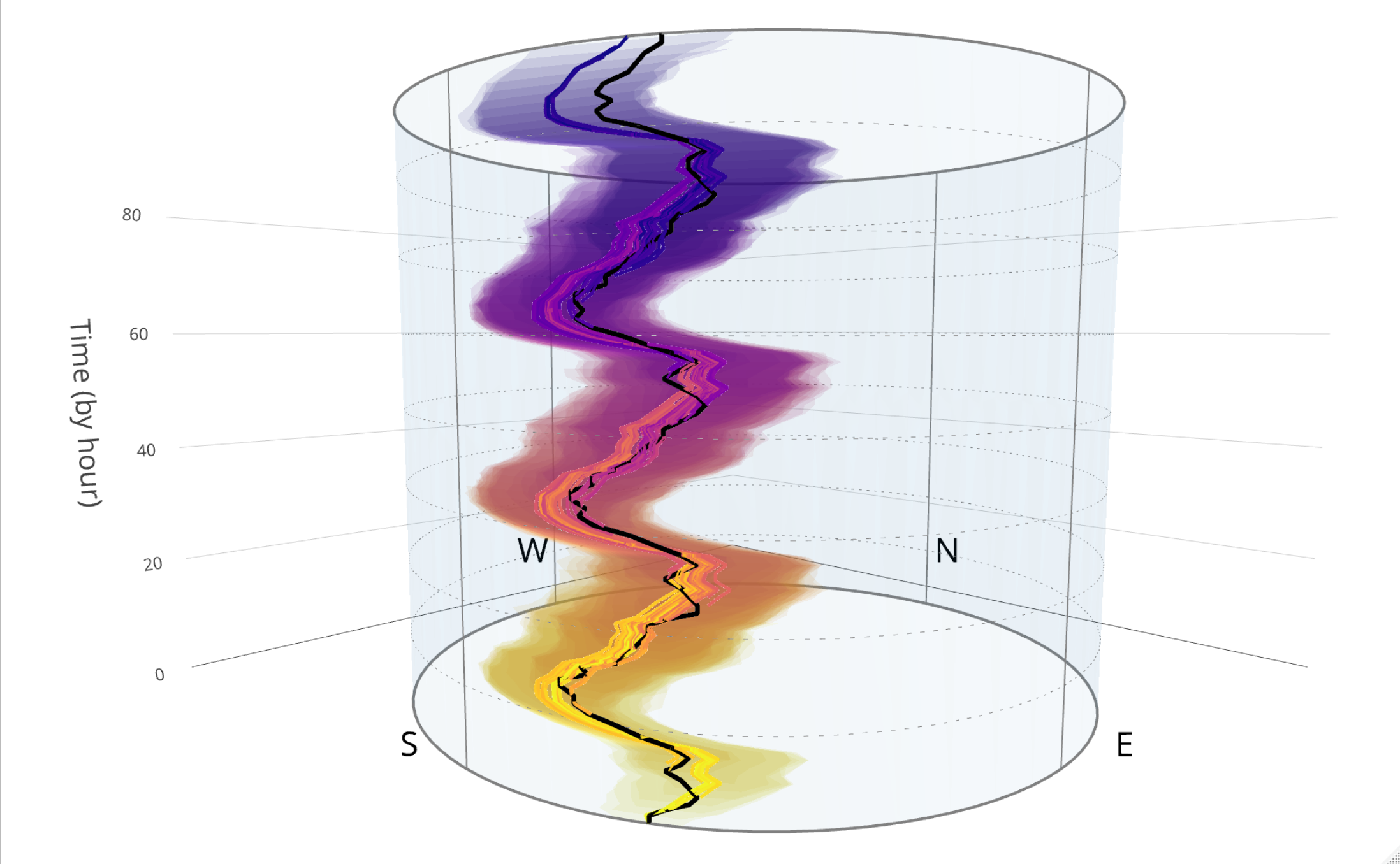}
  \end{minipage}
  \caption{Posterior predictive distributions for the 24-step ahead forecasting model. The left panel shows the predictive distributions of wind direction ($x$) and wind speed ($y$). The right panel presents a cylindrical representation of the predictive distribution for wind direction, where height corresponds to time (in hours) and angular position represents direction. Black points and curves denote the observed data, whilst colored curves and shaded regions represent posterior predictive means and 95\% credible intervals. Different colors correspond to forecasts starting from different time points.}
  \label{fig:wind_forecast}
\end{figure}

As an illustration, a 24-hour-ahead multi-step forecasting model is constructed using data observed in November and December. The 24-step forecasts are generated on a rolling basis, producing one 24-step forecast at each hour from 00:00 on December 28 to 23:00 on December 30, 2010 (72 different forecasting sequences). Assuming that only temperature information is observed, the forecasts for both wind direction and wind speed are obtained. The posterior predictive means and 95\% credible intervals at each forecast time point are shown in \autoref{fig:wind_forecast}.

From the plots, we observe that the forecast credible intervals cover the observed values and capture the daily patterns. The predictive means align well with the observed data in the first few steps, with larger errors as the forecasting step grows. The proposed circular regression framework enables efficient forecasting of multiple response variables simultaneously within a single model.

\subsection{Biomechanical data with joint circular and linear measurements}\label{sec5:biomechanical}

In this example, we use the biomechanical data from \citep{nagar2024dependent}, introduced in Section~\ref{sec1:joint_modeling}. The dataset consists of linear displacements in three directions ($y_{1}$, $y_{2}$, $y_{3}$) and three angular displacements, treated as circular variables ($x_{1}$, $x_{2}$, $x_{3}$). The dataset consists of 160 observations. The dependence among these displacement directions is studied using a six-dimensional joint Bayesian model given by:
\begin{equation}
    \label{eq:model_iid6d}
    \begin{aligned}
        x_{1i}\mid w_{1i}, \alpha_{1}, \kappa_{1} \sim \operatorname{LAvM}\left( \alpha_{1} + w_{1i}, \kappa_{1} \right)&, \quad y_{1i}\mid w_{4i}, \beta_{1},  \tau_{1} \sim N\left( \beta_{1} + w_{4i}, \tau_{1} \right), \\
        x_{2i}\mid w_{2i}, \alpha_{2}, \kappa_{2} \sim \operatorname{LAvM}\left( \alpha_{2} + w_{2i}, \kappa_{2} \right)&, \quad y_{2i}\mid w_{5i}, \beta_{2},  \tau_{2} \sim N\left( \beta_{2} + w_{5i}, \tau_{2} \right), \\
        x_{3i}\mid w_{3i}, \alpha_{3}, \kappa_{3} \sim \operatorname{LAvM}\left( \alpha_{3} + w_{3i}, \kappa_{3} \right)&, \quad y_{3i}\mid w_{6i}, \beta_{3},  \tau_{3} \sim N\left( \beta_{3} + w_{6i}, \tau_{3} \right), \\
        \mathbf{w}_{i} \sim N_{6}\left(0, \boldsymbol{\sigma}^{T}\mathbf{R}\boldsymbol{\sigma}\right), \quad \alpha_{1}, \alpha_{2}, \alpha_{3}, \beta_{1}&, \beta_{2}, \beta_{3} \sim N\left(0, 1\right), \quad \mathbf{R}\sim \operatorname{LKJ}\left(5\right),\\
        \sigma_{1},\ldots,\sigma_{6 } \sim \mathcal{PC}_{\tau}\left(1, 0.5\right)&, \quad \kappa_{1}, \kappa_{2}, \kappa_{3}, \tau_{1}, \tau_{2}, \tau_{3} = \exp\left\{15\right\}
    \end{aligned}
\end{equation}
where $\mathbf{w}_{i} = \left( w_{1i}, w_{2i}, w_{3i}, w_{4i}, w_{5i}, w_{6i} \right)^{T}$ is an i.i.d.\ six-dimensional latent Gaussian process with covariance matrix $\boldsymbol{\Sigma} = \boldsymbol{\sigma}^{T}\mathbf{R}\boldsymbol{\sigma}$. Here, $\boldsymbol{\sigma} = \left( \sigma_{1}, \sigma_{2}, \sigma_{3}, \sigma_{4}, \sigma_{5}, \sigma_{6} \right)^{T}$ is the vector of marginal standard deviations of the six model components and $\mathbf{R}$ is the correlation matrix with a Lewandowski-Kurowicka-Joe (LKJ) prior \citep{lewandowski2009generating} shared across all $i$.

In this model, we assume that the biomechanical instrument measures displacements with high accuracy. Accordingly, the concentration parameters ($\kappa_{1}$, $\kappa_{2}$, $\kappa_{3}$) and precision parameters ($\tau_{1}$, $\tau_{2}$, $\tau_{3}$) are fixed at large values, implying that only a small proportion of the variability is attributed to observational noise.  Under this specification, most of the structure and variability in the data are captured by the i.i.d. six-dimensional latent Gaussian process, rather than leaving it to an unstructured error term. Consequently, this modeling choice makes it possible to study dependence between different components and observations through the shared latent structure.
\begin{figure}[!ht]
  \centering
  \begin{minipage}[b]{0.5\textwidth}
    \includegraphics[width=\textwidth]{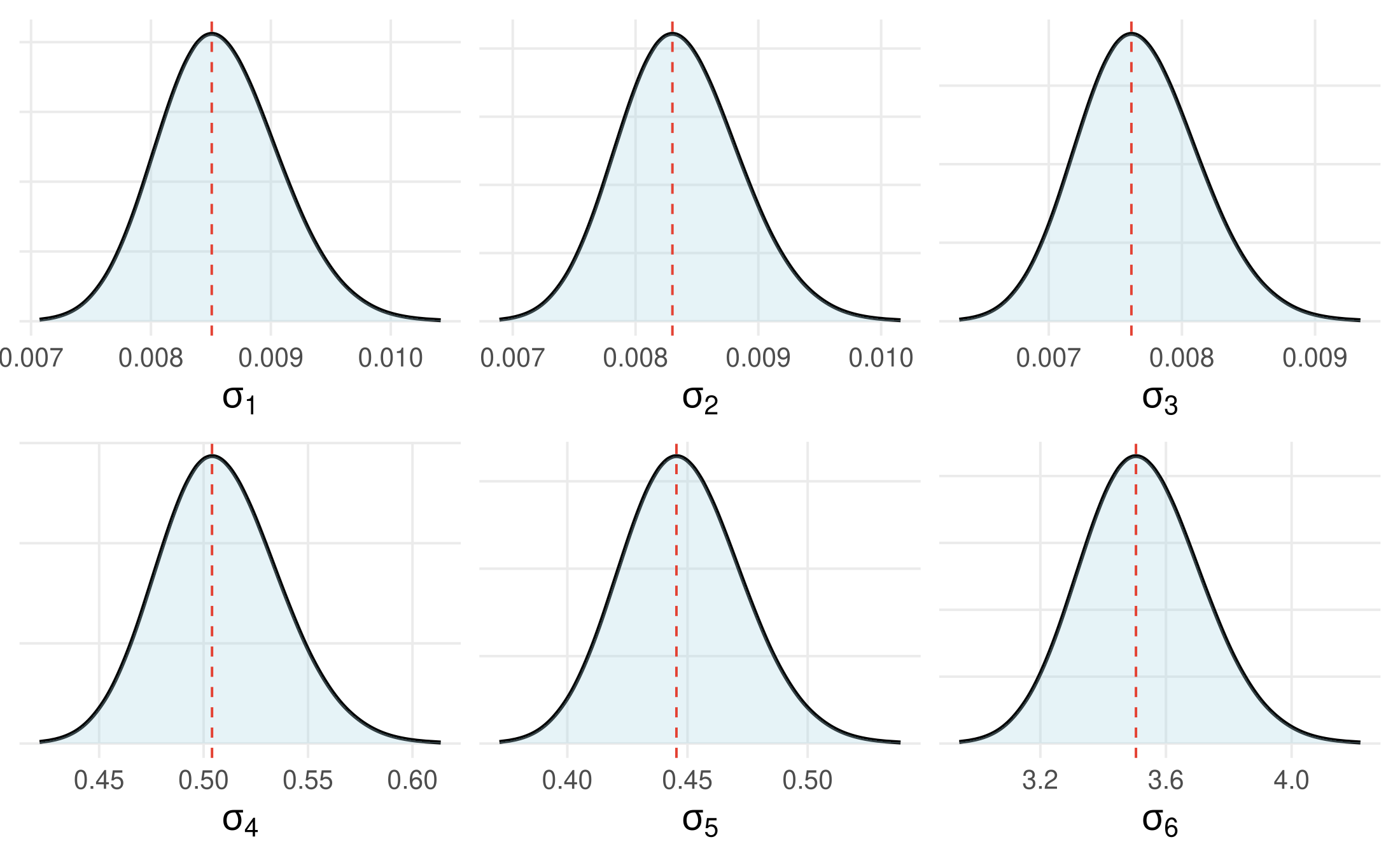}
  \end{minipage}
  \hfill
  \begin{minipage}[b]{0.49\textwidth}
    \includegraphics[width=\textwidth]{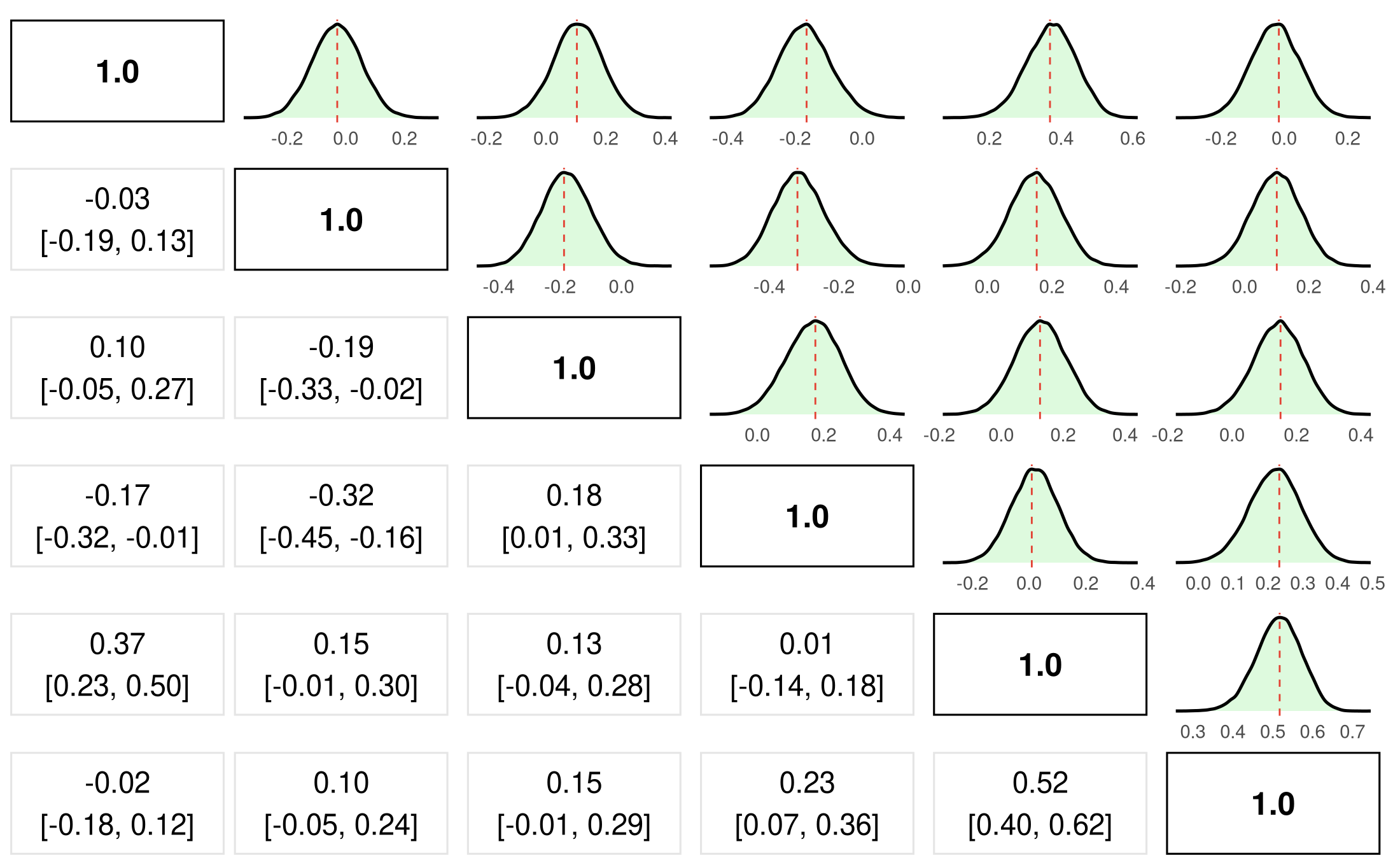}
  \end{minipage}
  \caption{Posterior marginals of the standard deviations (left) and the correlation matrix (right) of the six-dimensional latent Gaussian process in \autoref{eq:model_iid6d}. The black curves show posterior densities and red vertical lines indicate posterior modes.}
  \label{fig:bio}
\end{figure}

The running time of this model is only around 1 second. The posterior marginals of the intercepts and the posterior predictive distributions are provided in Appendix~\ref{appendix2}. The posterior mode of the covariance matrix $\mathbf{\Sigma}=\boldsymbol{\sigma}^{T}\mathbf{R}\boldsymbol{\sigma}$ for the six-dimensional latent Gaussian process is given by
\begin{equation}
    \label{eq:correlation_bio}
    \begin{aligned}
    \boldsymbol{\sigma} &= \scalebox{0.75}{$\begin{bmatrix} 0.0085 \\ 0.0083 \\ 0.0076 \\ 0.5056 \\ 0.4468 \\ 3.5145 \end{bmatrix}$}, \qquad
    \mathbf{R}  &= \scalebox{0.75}{$\begin{bmatrix}
    1.00 & -0.03 & 0.11 & -0.17 & \textbf{0.37} & -0.03 \\
    -0.03 & 1.00 & -0.18 & \textbf{-0.31} & 0.15 & 0.10 \\
    0.11 & -0.18 & 1.00 & 0.17 & 0.13 & 0.14 \\
    -0.17 & -0.31 & 0.17 & 1.00 & 0.02 & 0.22 \\
    0.37 & 0.15 & 0.13 & 0.02 & 1.00 & \textbf{0.52} \\
    -0.03 & 0.10 & 0.14 & 0.22 & 0.52 & 1.00
    \end{bmatrix}$}.
    \end{aligned}
\end{equation}
The posterior marginals of $\boldsymbol{\sigma}$ and $\mathbf{R}$ are shown in \autoref{fig:bio}. From $\mathbf{R}$, we observe that the linear movement $y_{2}$ and $y_{3}$ are strongly correlated with each other, suggesting coupled motion in these two directions. In addition, the pairs $\left(x_{1}, y_{2}\right)$ and $\left(x_{2},y_{1}\right)$ show non-negligible correlations, indicating interactions between angular and linear displacements. The identified patterns suggest that stabilizing angular motion along the $x_{1}$ and $x_{2}$ angles, as well as jointly controlling movement in the $y_{2}$ and $y_{3}$ axes, may help reduce overall linear displacement and improve fixator stability.


\section{Discussion}\label{sec6}

The cyclic nature of geodesic distance can lead to inferential ambiguity in the estimation of circular statistical models, whilst the incompatibility between circular spaces and linear spaces complicates the joint analysis of circular and linear variables. In this paper, we introduce a regression-oriented adjustment of circular distributions that addresses the inferential ambiguity issue. We have developed a Bayesian regression framework that allows the information in circular variables to be incorporated either as responses or as components of the regression predictor, across a wide range of model settings.

The proposed framework is broadly applicable within Bayesian modeling, and can achieve computationally highly efficient inference when implemented using INLA, owing to the multi-layer structure of the model and INLA's strengths in handling latent Gaussian processes. The model aligns naturally with the INLA framework, making it both efficient and flexible for constructing models with structured dependence, such as spatial and spatio-temporal models, including settings with additional covariates.

A limitation of the proposed framework is that it is not well suited to situations in which the circular data are close to uniformly distributed. In such cases, the LAvM or other LAC densities are not appropriate, as the adjustment relies on the circular response being reasonably concentrated.

The proposed distributions, priors, and modeling framework are currently being implemented in a user-friendly R package to facilitate practical use.

\newpage


\begin{appendices}
\section{Derivation of Definition \ref{def:LAC}}\label{appendix1}

The derivation for the LAC density in Definition~\ref{def:LAC} follows the argument given below with minor modifications:

Let $Z \sim \operatorname{vM}\left(0, \kappa\right)$, then its probability density function is
\begin{align}
    p_{Z}\left(z \mid \kappa\right) &= \frac{1}{2\pi I_{0}\left(\kappa\right)} \exp \left\{ \kappa \cos \left( z \right) \right\}, \qquad z \in \left[-\pi,\pi\right).
\end{align}
Let circular variable $X\in\left[-\pi,\pi\right)$ satisfies
\begin{align}
    Z := h\left(X\right) = g\left(g^{-1}\left(X\right) - \eta\right) \Longleftrightarrow g^{-1}\left(Z\right) = g^{-1}\left(X\right) - \eta
\end{align}
where $g^{-1}\left(\cdot\right): \mathcal{S} \rightarrow \mathbb{R}$ and $g\left(\cdot\right): \mathbb{R} \rightarrow \mathcal{S}$, so $h\left(\cdot\right): \mathcal{S}\rightarrow\mathcal{S}$.

Let $\left\lvert \frac{dZ}{dx} \right\rvert$ denote the Jacobian for the change of variables, then by letting $u = g^{-1}\left(X\right) - \eta$, we have
\begin{equation}
    \begin{aligned}
        \frac{dZ}{dx} &= \frac{d g\left(u\right)}{d u} \frac{d u}{dx} = \frac{1}{\frac{d \, g^{-1}\left(g\left(u\right)\right)}{d \, \left(g\left(u\right)\right)}}\frac{d \, \left(g^{-1}\left(x\right) - \eta\right)}{d \, x} = \frac{\left(g^{-1}\right)'\left(x\right)}{\left(g^{-1}\right)'\left(z\right)} .
    \end{aligned}
\end{equation}
Then, the probability density function for $X$, that is, the density for LAvM, is
\begin{align}
    p_{\operatorname{LAvM}}\left(x \mid \eta, \kappa\right) &= p_{Z}\left(h\left(x\right)\right) \left\lvert \frac{dZ}{dx} \right\rvert = \frac{\exp \left\{ \kappa \cos \left( g\left(g^{-1}\left(x\right) - \eta\right) \right) \right\}}{2\pi I_{0}\left(\kappa\right)} \left\lvert \frac{\left(g^{-1}\right)'\left(x\right)}{\left(g^{-1}\right)'\left(z\right)} \right\rvert,
\end{align}
where $z = g\left(g^{-1}\left(x\right) - \eta\right)$.

When $g\left(y\right) = 2\arctan\left(y\right)$ and $g^{-1}\left(y\right) = \tan\left(\frac{y}{2}\right)$, then $z = 2 \arctan \left( \tan \left( \frac{x}{2} \right) - \eta \right)$, and $\left(g^{-1}\right)'\left(z\right) = \frac{1}{2\cos^{2}\left(z/2\right)}$. The corresponding LAvM density for $X$ is
\begin{equation}
    \begin{aligned}
        p_{\operatorname{LAvM}}&\left(x \mid \eta, \kappa\right) = p_{Z}\left(h\left(x\right)\right) \cdot \left\lvert \frac{\left(g^{-1}\right)'\left(x\right)}{\left(g^{-1}\right)'\left(z\right)} \right\rvert = \frac{\exp\left\{ \kappa \cos\left( z \right) \right\}}{2\pi I_{0}\left(\kappa\right)}\cdot \frac{\cos^{2}\left(z/2\right)}{\cos^{2}\left(x/2\right)} \\
        &= \frac{\exp\left\{ \kappa \cos\left( 2 \arctan \left( \tan \left( \frac{x}{2} \right) - \eta \right) \right) \right\}}{2\pi I_{0}\left(\kappa\right)\left(1 + \eta^{2} - \eta \sin\left(x\right) - \eta^{2}\sin^{2}\left(\frac{x}{2}\right)\right)} .
    \end{aligned}
\end{equation}

\subsection{Concentration of LAvM distribution} \label{appendix1:proof1_property}

The Gaussian approximation \citep{tierney1986accurate} of LAvM is $N\left( x^{*}, -\frac{1}{\ell^{''}\left(x^{*}\right)} \right)$ where $x^{*}$ is the mode for LAvM, and $\ell^{''}\left(x^{*}\right)$ is the second derivation the log-likelihood of $x$ evaluated at the mode.

For simplicity, let $h\left(\cdot\right) = g^{-1}\left(\cdot\right)$ and $z'\left(x\right) = \frac{dz}{dx} = \frac{h'\left(x\right)}{h'\left(z\right)}$, then the log-density of LAvM is
\begin{align}
    \ell(x) = \kappa \cos\left(z\right) + \log \left\lvert h'\left(x\right)\right\rvert - \log \left\lvert h'\left(z\right)\right\rvert.
\end{align}
Let $S\left(y\right) = \frac{d}{dy} \log \left\lvert h'\left(y\right)\right\rvert = \frac{h^{''}\left(y\right)}{h'\left(y\right)}$, then
\begin{equation}
    \begin{aligned}
    \ell'\left(x\right) &= \frac{d}{dx} \left( \kappa \cos(z) \right) + S\left(x\right) - \frac{d}{dx} \log \left\lvert h'\left(z\right)\right\rvert\\
    &= -\kappa \sin\left(z\right) \cdot z'\left(x\right)\left(x\right) + S\left(x\right) - S\left(z\right) \cdot z'\left(x\right) \\
    &= z'\left(x\right) \left( -\kappa \sin\left(z\right) - S\left(z\right) \right) + S\left(x\right).
    \end{aligned}
\end{equation}
Now, we find the derivative of $S\left(y\right)$ and $z'\left(x\right)$:
\begin{align}
    Q\left(y\right) &= S'\left(y\right) = \frac{d}{dy} \left( \frac{h^{''}\left(y\right)}{h'\left(y\right)} \right) = \frac{h^{'''}\left(y\right)h'\left(y\right) - \left(h^{''}\left(y\right)\right)^2}{\left(h'\left(y\right)\right)^2} \\
    z^{''}\left(x\right) &= \frac{d}{dx} \left( \frac{h'\left(x\right)}{h'\left(z\right)} \right) = z'\left(x\right) \left( S\left(x\right) - z'\left(x\right)S\left(z\right) \right).
\end{align}
Therefore,
\begin{equation}
    \begin{aligned}
        \ell^{''}\left(x\right) &= z^{''}\left(x\right) \left[ -\kappa \sin(z) - S(z) \right] + z'\left(x\right) \frac{d}{dx} \left[ -\kappa \sin(z) - S(z) \right] + Q(x) \\ &= z^{''}\left(x\right) \left[ -\kappa \sin(z) - S(z) \right] + z'\left(x\right) \left[ -\kappa \cos(z) \cdot z'\left(x\right) - Q(z) \cdot z'\left(x\right) \right] + Q(x) \\ &= z^{''}\left(x\right) \left[ -\kappa \sin(z) - S(z) \right] - (z'\left(x\right))^2 \left[ \kappa \cos(z) + Q(z) \right] + Q(x).
    \end{aligned}
\end{equation}
Since the mode is $h\left(x^{*}\right)-\eta = 0 \Longleftrightarrow x^{*} = g\left(\eta\right)$, we have $z\left(x^{*}\right) = g\left(0\right) = 0$. Therefore, $z'\left(x^{*}\right) = h'\left(x^{*}\right)/h'\left(0\right)$, and
\begin{align}
    z^{''}\left(x^{*}\right) &= z'\left(x^{*}\right) \left[ S\left(x^{*}\right) - z'\left(x^{*}\right) S\left(0\right) \right].
\end{align}
Note that $\sin\left(0\right)=0$ and $\cos\left(0\right)=1$, thus,
\begin{equation}
    \begin{aligned}
        \ell^{''}\left(x^{*}\right) &= z^{''}\left(x^{*}\right) \left[ - S\left(0\right) \right] - \left(z'\left(x^{*}\right)\right)^2 \left[ \kappa + Q\left(0\right) \right] + Q\left(x^{*}\right) \\
        &= -S\left(0\right) \cdot z'\left(x^{*}\right) \left[ S\left(x^{*}\right) - z'\left(x^{*}\right) S\left(0\right) \right] - \left(z'\left(x^{*}\right)\right)^2 \left[ \kappa + Q\left(0\right) \right] + Q\left(x^{*}\right) \\
        &= -z'\left(x^{*}\right) S\left(x^{*}\right) S\left(0\right) + \left(z'\left(x^{*}\right)\right)^2 S\left(0\right)^2 \\
        &- \left(z'\left(x^{*}\right)\right)^2 \kappa - \left(z'\left(x^{*}\right)\right)^2 Q\left(0\right) + Q\left(x^{*}\right) \\
        &= Q\left(x^{*}\right) - \left(z'\left(x^{*}\right)\right)^2 \left[ \kappa + Q\left(0\right) - S\left(0\right)^2 \right] - z'\left(x^{*}\right) S\left(x^{*}\right) S\left(0\right).
    \end{aligned}
\end{equation}
Hence, the precision for the Gaussian approximation of LAvM distribution is
\begin{align}
    \label{eq:LAvM_precision}
    \tau = -\ell^{''}\left(x^{*}\right) = \left(z'\left(x^{*}\right)\right)^2 \left[ \kappa + Q\left(0\right) - S\left(0\right)^2 \right] + z'\left(x^{*}\right) S\left(x^{*}\right) S\left(0\right) - Q\left(x^{*}\right)
\end{align}
Recall that the Gaussian approximation of the von Mises distribution have precision $\kappa$, which equals to the concentration parameter of von Mises distribution. From \autoref{eq:LAvM_precision}, we observe that the approximate concentration parameter for LAvM density depends on $x^{*}=g\left(\eta\right)$, and thus depends on the linear predictor $\eta$.

When $g\left(y\right) = 2\arctan\left(y\right)$ and $h\left(y\right) = \tan\left(\frac{y}{2}\right)$, recall the notations and derivations in Appendix~\ref{appendix1:proof1_property}, we have
\begin{align}
    h'\left(y\right) &= \frac{1}{2\cos^{2}\left(y/2\right)} = \frac{1}{2}\sec^{2}\left(\frac{y}{2}\right) = \frac{1}{2}\left(1 + \tan^{2}\left(\frac{y}{2}\right)\right), \\
    \Longrightarrow h^{''}\left(y\right) &= \frac{d}{dy}\left[ \frac{1}{2}\left(1 + h\left(y\right)^{2}\right) \right] = \frac{1}{2} \cdot 2 h\left(y\right) \cdot h'\left(y\right) = h\left(y\right)h'\left(y\right).
\end{align}
At the mode, $\tan\left(x^{*}/2\right) = \eta \Longleftrightarrow x^{*}=2\arctan\left(\eta\right)$, thus
\begin{align}
    z'\left(x^{*}\right) = \frac{h'\left(x^{*}\right)}{h'\left(0\right)} = \frac{\frac{1}{2}\left(1+\eta^{2}\right)}{\frac{1}{2}} = 1 + \eta^{2}
\end{align}
Now derive $S\left(y\right)$ and $Q\left(y\right)$:
\begin{align}
    S\left(y\right) &= \frac{h^{''}\left(y\right)}{h'\left(y\right)} = \frac{h\left(y\right)h'\left(y\right)}{h'\left(y\right)} = h\left(y\right) = \tan\left(\frac{y}{2}\right), \\
    \Longrightarrow Q\left(y\right) &= S'\left(y\right) = \frac{1}{2}\left(1 + \tan^{2}\left(\frac{y}{2}\right)\right).
\end{align}
Therefore, $S\left(0\right) = 0$, $Q\left(0\right) = \frac{1}{2}$, $S\left(x^{*}\right) = \eta$ and $Q\left(x^{*}\right) = \frac{1}{2}\left(1 + \eta^{2}\right)$. Recall \autoref{eq:LAvM_precision}, the precision of the Gaussian approximation of the LAvM distribution with inverse tangent link is
\begin{equation}
    \begin{aligned}
        \tau &= \left(z'\left(x^{*}\right)\right)^2 \left[ \kappa + Q\left(0\right) - S\left(0\right)^2 \right] + z'\left(x^{*}\right) S\left(x^{*}\right) S\left(0\right) - Q\left(x^{*}\right) \\
        &= \left(1 + \eta^{2}\right)^{2}\left( \frac{1}{2} + \kappa \right) - \frac{1}{2}\left(1 + \eta^{2}\right) = \kappa\left(1 + \eta^{2}\right)^{2} + \frac{1}{2}\eta^{2}\left(1 + \eta^{2}\right).
    \end{aligned}
\end{equation}
Hence, the approximate concentration parameter of LAvM distribution with inverse tangent link is $\kappa\left(1 + \eta^{2}\right)^{2} + \frac{1}{2}\eta^{2}\left(1 + \eta^{2}\right)$. When $\kappa \gg \eta$, then $\kappa\left(1 + \eta^{2}\right)^{2} + \frac{1}{2}\eta^{2}\left(1 + \eta^{2}\right) \approx \kappa\left(1 + \eta^{2}\right)^{2}$.

\newpage

\section{Full results for simulation studies and application examples}\label{appendix2}

\begin{figure}[!ht]
    \centering\includegraphics[width=0.9\linewidth]{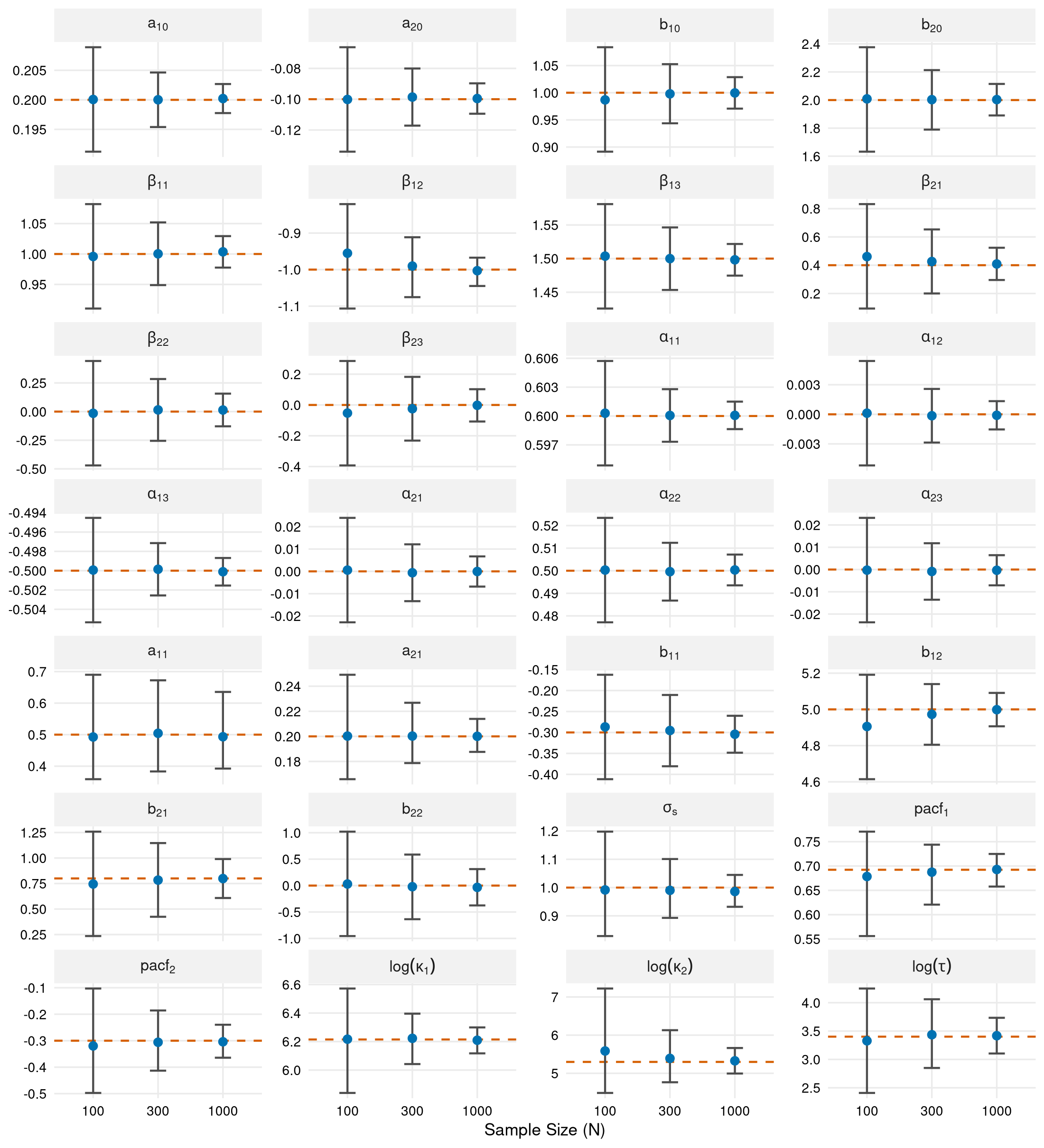}
    \caption{Posterior summaries for the simulation in Section~\ref{sec4:complex_model}. The same plotting convention as in \autoref{fig:sim1} is used. The blue dots denote posterior means for the parameters shown in the first four rows and posterior modes for the remaining parameters.}
    \label{fig:res_sim3}
\end{figure}

\begin{figure}[!ht]
    \centering\includegraphics[width=0.9\linewidth]{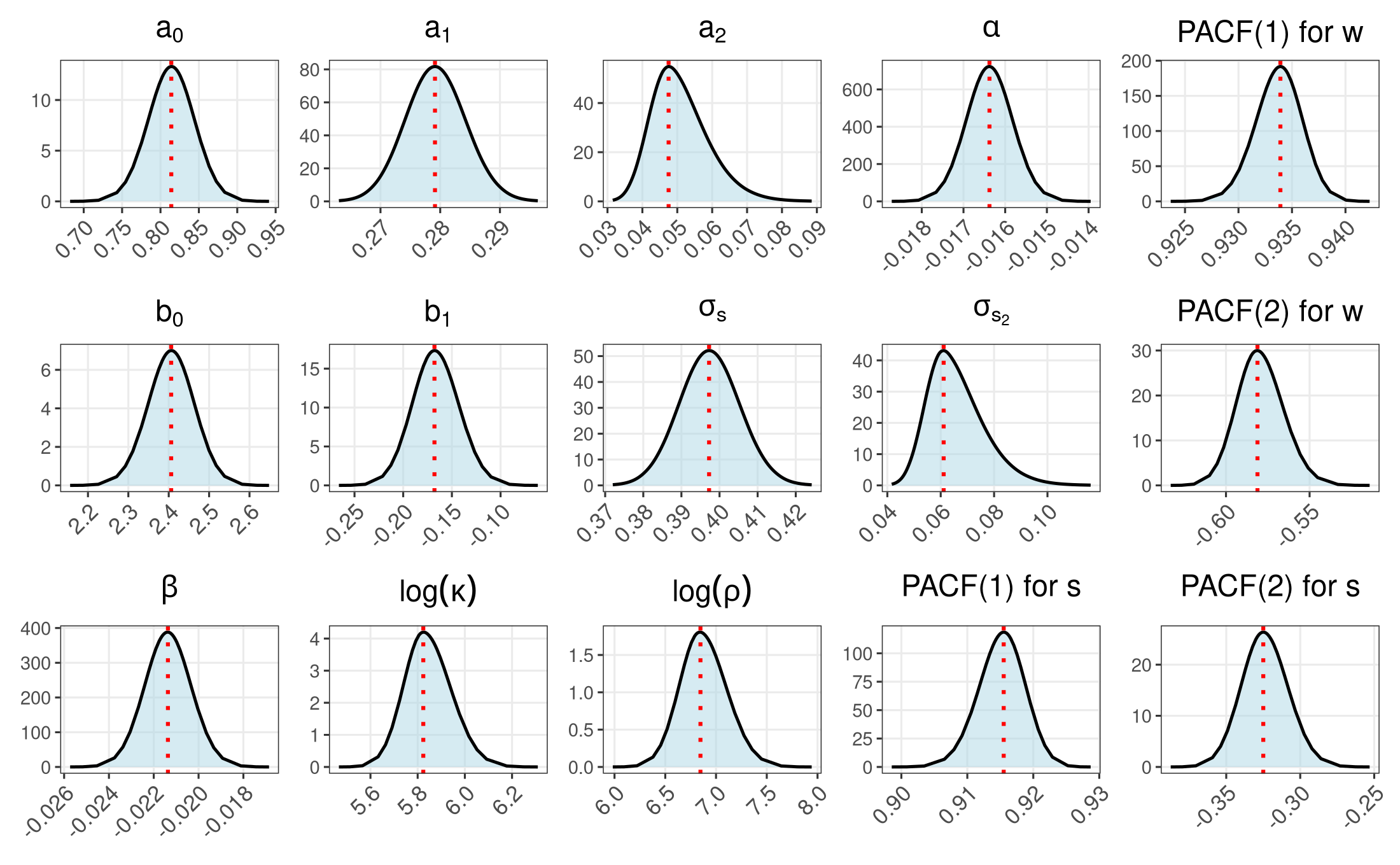}
    \caption{Posterior summaries for the wind model in Section~\ref{sec5:wind}.}
    \label{fig:wind.posteriors}
\end{figure}

\begin{figure}[!ht]
  \centering
  \begin{minipage}[b]{0.5\textwidth}
    \includegraphics[width=\textwidth]{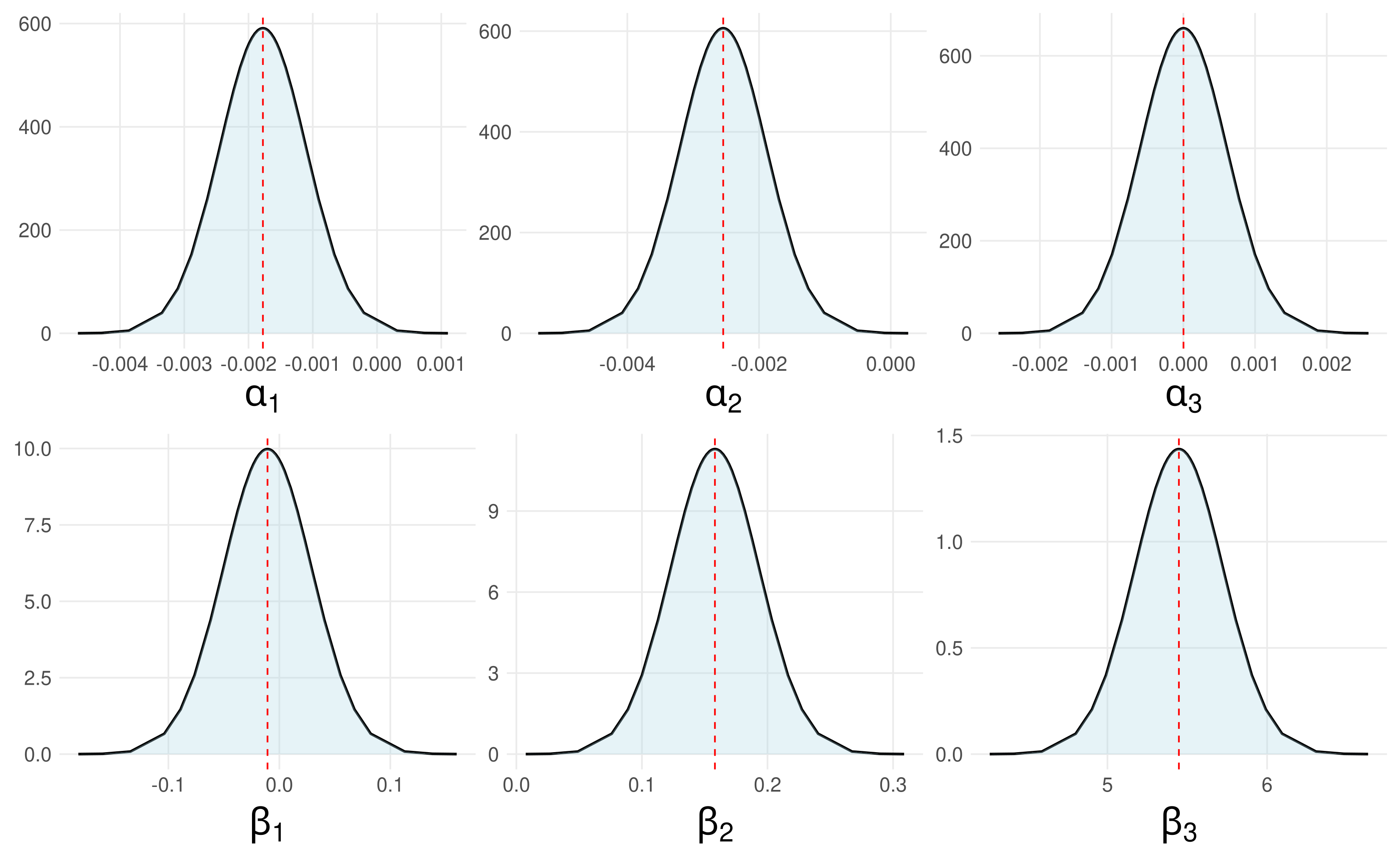}
  \end{minipage}
  \hfill
  \begin{minipage}[b]{0.49\textwidth}
    \includegraphics[width=\textwidth]{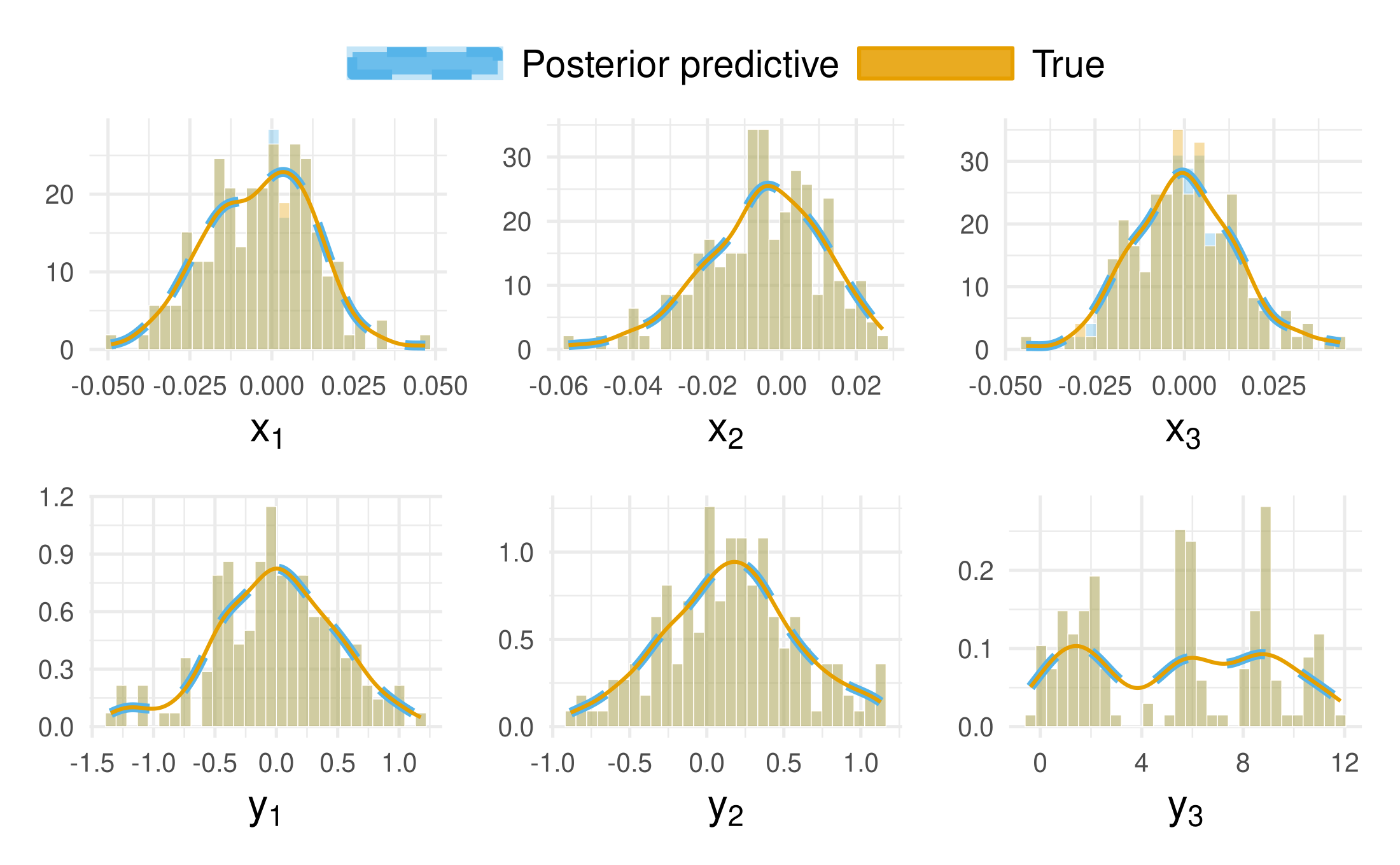}
  \end{minipage}
  \caption{Posterior marginals of the intercepts (left) and posterior predictive distributions (right) for the biomechanical model in \autoref{eq:model_iid6d}. In the left panel, black curves show posterior densities and red vertical lines indicate posterior means. The posterior predictive plots follow the same convention as in \autoref{fig:sim1}.}
  \label{fig:bio_appendix}
\end{figure}
\end{appendices}

\bibliographystyle{abbrvnat}
\bibliography{references}

\end{document}